\documentclass[3p,10pt]{elsarticle}

\journal{Information Sciences}

\usepackage{lineno}

\usepackage{hyperref}
\makeatletter
\g@addto@macro{\UrlBreaks}{\UrlOrds}
\makeatother

\usepackage[dvipsnames]{xcolor}

\usepackage{adjustbox}

\usepackage{hyperref}
\usepackage{amsmath}
\usepackage{amsfonts}
\usepackage[algo2e,ruled,lined]{algorithm2e} 

\usepackage{ntheorem}

\newtheorem{myprob}{Problem}

\newtheorem{mydef}{Definition}

\usepackage{bbm}
\usepackage{dsfont}

\usepackage{algorithm}[h]
\usepackage[noend]{algpseudocode}

\usepackage{paralist}
\usepackage{booktabs} %
\usepackage{multirow}
\usepackage{subfigure}

\usepackage[labelfont=bf,belowskip=0pt,aboveskip=3pt]{caption}

\usepackage{graphicx}
\usepackage{epstopdf}
\usepackage{cancel}
\usepackage{soul}\DeclareGraphicsExtensions{.eps,.gz,.eps,.pdf,.png,.jpg}
\usepackage{ulem}

\graphicspath{{./}{./fig/}}

\newcommand{\sstitle}[1]{\smallskip\noindent\textbf{#1.\/}}

\def\Snospace~{\S{}}

\newcommand{\romanNum}[1]{\lowercase\expandafter{\romannumeral #1\relax}}

\makeatletter
\newcommand{\removelatexerror}{\let\@latex@error\@gobble}
\makeatother

\begin{document}

\setlength{\belowdisplayskip}{3pt}
\setlength{\belowdisplayshortskip}{3pt}
\setlength{\abovedisplayskip}{3pt}
\setlength{\abovedisplayshortskip}{3pt}

\begin{frontmatter}

\title{Heterogeneous Hypergraph Embedding for Recommendation Systems}

\author[griffith]{Darnbi Sakong\fnref{note1}}
\author[hust]{Viet Hung Vu\fnref{note1}}
\fntext[note1]{The first two authors have equal contribution}
\author[epfl]{Thanh Trung Huynh}
\author[hust]{Phi Le Nguyen}

\author[uq]{Hongzhi Yin}
\author[griffith]{Quoc Viet Hung Nguyen}
\author[griffith]{Thanh Tam Nguyen\corref{cor1}}
\cortext[cor1]{Corresponding author}

\affiliation[griffith]{organization={Griffith University},
                country={Australia}}

\affiliation[hust]{organization={Hanoi University of Science and Technology},
                country={Vietnam}}

\affiliation[epfl]{organization={Ecole Polytechnique Federale de Lausanne},
                country={Switzerland}}
                
\affiliation[uq]{organization={The University of Queensland},
                country={Australia}}

\begin{abstract}
Recent advancements in recommender systems have focused on integrating knowledge graphs (KGs) to leverage their auxiliary information. The core idea of KG-enhanced recommenders is to incorporate rich semantic information for more accurate recommendations. However, two main challenges persist: i) Neglecting complex higher-order interactions in the KG-based user-item network, potentially leading to sub-optimal recommendations, and ii) Dealing with the heterogeneous modalities of input sources, such as user-item bipartite graphs and KGs, which may introduce noise and inaccuracies. To address these issues, we present a novel Knowledge-enhanced Heterogeneous Hypergraph Recommender System (KHGRec). KHGRec captures group-wise characteristics of both the interaction network and the KG, modeling complex connections in the KG. Using a collaborative knowledge heterogeneous hypergraph (CKHG), it employs two hypergraph encoders to model group-wise interdependencies and ensure explainability. Additionally, it fuses signals from the input graphs with cross-view self-supervised learning and attention mechanisms. Extensive experiments on four real-world datasets show our model's superiority over various state-of-the-art baselines, with an average 5.18\% relative improvement. Additional tests on noise resilience, missing data, and cold-start problems demonstrate the robustness of our KHGRec framework. Our model and evaluation datasets are publicly available at \url{https://github.com/viethungvu1998/KHGRec}.
\end{abstract}

\begin{keyword}
hypergraph embedding, knowledge-based recommender systems, self-supervised learning, graph-based collaborative filtering 
\end{keyword}

\end{frontmatter}

%\linenumbers

\section{Introduction}
\label{sec:intro}
Recommender Systems (RecSys) provide personalized suggestions to users by collecting and analyzing their past preferences and behaviors such as user profiles and user-item interactions~\cite{nguyen2024csur,perifanis2023fedpoirec,xu2023empowering,liu2024matrix}. RecSys plays an essential role in various applications such as e-commerce websites, streaming services, social media platforms, news portals, e-healthcare~\cite{pham2023hierarchical}. In these applications, collaborative filtering (CF) methods are frequently employed, where items are recommended by the likes or actions of users with similar tastes. With the recent progression in deep learning structures, notably graph neural networks~\cite{shu2023metagc}, various strategies integrating graph to conventional CF methods have emerged~\cite{he2020lightgcn,xia2022hypergraph}. These strategies effectively capture intricate connections between users and items, offering a comprehensive perspective on the interaction data.

Despite these advancements, most existing CF techniques rely solely on user-item interaction data. The inherent limitation due to the scarcity of these interactions fundamentally restricts further enhancement in performance. In response to this challenge, integrating knowledge graph (KG)~\cite{li2023efficient} has emerged as a prominent strategy in collaborative filtering, referred to as \textit{KG-enhanced collaborative filtering RecSys}. Such techniques provide a comprehensive information network for items, consequently resulting in recommendations that are enhanced by the knowledge graph (KG)~\cite{zhang2018learning, yang2018knowledge}. For example, CKE \cite{zhang2016collaborative} incorporates both topological and multi-modal information to generate embeddings. KGAT~\cite{wang2019kgat} introduces a mechanism to combine both user-item interaction and knowledge signal to construct a unified relational graph. This line of research opens a new direction for personalized Recsys by providing users with contextually relevant recommendations that match the user's preferences. 

These models leverage the multiple layers aggregation mechanism to consider high-order interactions to an extent. Nevertheless, the current knowledge-enhanced collaborative filtering RecSys models built upon classical graph structures, where each edge primarily connects a pair of nodes, have proven inadequate in capturing the essential higher-order characteristics intrinsic to knowledge graphs (KGs)~\cite{poskhg}. In real-world applications, the nature of relationships among objects frequently extends beyond simple pairwise interactions to include triadic, tetradic, or more complex configurations. Simplifying these complex relationships into binary ones results in a loss of critical information and reduces the system's expressiveness. Consequently, it limits the quality of recommendations as it neglects to identify user item groups sharing common patterns. For example, in movie recommendation systems like MovieLens and Netflix, users typically derive satisfaction from films within their preferred genre (e.g., action, horror, romance) or those featuring their beloved actors (as depicted in Fig.\ref{fig:case_study}). This highlights the need for integrating group-wise (a.k.a higher-order) interaction within the recommendation process.

The integration of external data sources such as KG also raises the challenge of heterogeneous modality to the RecSys. Prior research~\cite{wang2019kgat} involves dynamic item embedding updates across the user-item and the external knowledge graph, capturing signals from diverse data sources but potentially introducing sub-optimal parameters due to noise. Recent methods like KGRec~\cite{kgrec} independently learn user-item and knowledge graphs for distinct item embeddings, while KGCL~\cite{kgcl} employs KG-enhanced item representations to guide cross-view contrastive learning, reducing noise. Despite representing the same item nodes, these embeddings exhibit varied distributions due to differing graph connectivity. Thus, the integration of data with different modalities requires careful consideration for possible conflict and noises. 

We argue that a knowledge-assisted recommender system framework should be capable of handling the following challenges: 
\begin{itemize}
    \item \textbf{C1: Group-wise dynamics:} The collaborative and knowledge graph has group-wise characteristics, e.g., a user interacts with multiple items, or an item relates to multiple entities. A solution thus needs to be capable of efficiently representing such characteristics.     
    \item \textbf{C2: KG relational dependency:} The relations between entities in KG are very complex. For instance, a film could be connected to various other entities through different relations - it could be directed by one entity (a director), starring several others (actors), belong to a certain genre, and so on. Each of these connections represents a different type of relationship, and together, they form a rich and intricate network of relational dependencies. Therefore, a solution should be capable of analyzing and understanding these relational dependencies.
    \item \textbf{C3: Explainability:} In conventional collaborative filtering recommendation systems, suggestions are predominantly grounded in user-item interactions. This approach may yield recommendations that lack transparency and clarity. Consequently, an explainable knowledge-based recommender system is necessitated, one that is capable of providing insights into the rationale behind the selection of specific items. 
    \item \textbf{C4: Consistency:} In both collaborative and knowledge graphs, item entities may appear, creating a unique challenge. Specifically, latent features of the same entity from different latent embeddings should be close in proximity in the embedding space. This is based on the assumption that the same entity should have a similar representation across different graphs, reflecting its consistent characteristics. An effective solution, thus, should be capable of aligning entities of the same item across different embeddings.
\end{itemize}

To tackle the outlined challenges, this paper proposes a novel knowledge-based collaborative filtering Recsys named \textit{\underline{K}nowledge-enhanced \underline{H}eterogeneous Hyper\underline{g}raph \underline{Rec}ommender System} (KHGRec). Specifically, our method leverages heterogeneous hypergraph to better integrate the user-item bipartite graph and knowledge graph under the same modal as well as naturally capture the higher-order characteristics among the nodes. We then introduce a heterogeneous hypergraph convolution applied to the constructed hypergraph to simultaneously embed the group-wise characteristics of both input graphs and capture the complex relation-aware connections in the KG.

The contributions of this work are summarized as follows:
\begin{itemize}
    \item We propose KHGRec, a novel knowledge-enhanced collaborative filtering Recsys, which is the first method to couple the hypergraph's group-wise characteristics with the explainability of knowledge-enhanced Recsys to improve the recommendation's quality.
    \item Drawing from input data, which includes user-item bipartite and knowledge graphs, we present a method for constructing a hypergraph data structure known as the Collaborative Knowledge Heterogeneous Hypergraph. This structure unifies the input graphs without introducing noises and facilitates the later representation learning.
    
    \item We present a novel relational-aware hypergraph neural network, innovatively designed for mining the nodes' higher-order interactions and highlighting the node significance within hyperedges using attention mechanism, taking into account their relationships. 

    \item To seamlessly incorporate the signal retrieved from the two input graphs, we leverage the mechanism including two components. The first component leverages the attention mechanism to learn the optimal weight assigned for each latent matrix. Additionally, a cross-view self-supervised learning mechanism is employed to align similar entities from different latent spaces.
    \item We perform extensive experiments with nine baselines on two popular datasets to justify the model's performance against state-of-the-art models. Our source code and dataset are publicly available.~\footnote{https://github.com/viethungvu1998/KHGRec}
\end{itemize}

The remainder of the paper is organized as follows. \autoref{sec:problem} gives a motivating example, then introduces the problem statement and provides a comprehensive overview of our approach. \autoref{sec:constructor} describes the process of constructing the heterogeneous hypergraph from the user-item bipartite and the knowledge graph. \autoref{sec:embed} and \autoref{sec:contrastlearn} explain the two main components of our proposed framework. \autoref{sec:exp} reports the experiments we conducted to study the performance of our technique compared to the state-of-the-art baselines. \autoref{sec:related} reviews related work and \autoref{sec:con} concludes the paper.

\section{Problem and Approach}
\label{sec:problem} 

This section first introduces a motivating example and then describes the key structures and problem formulation for our paper.

\subsection{Motivating example}

\begin{figure}[!h]
	\centering
	\includegraphics[width=0.7\linewidth]{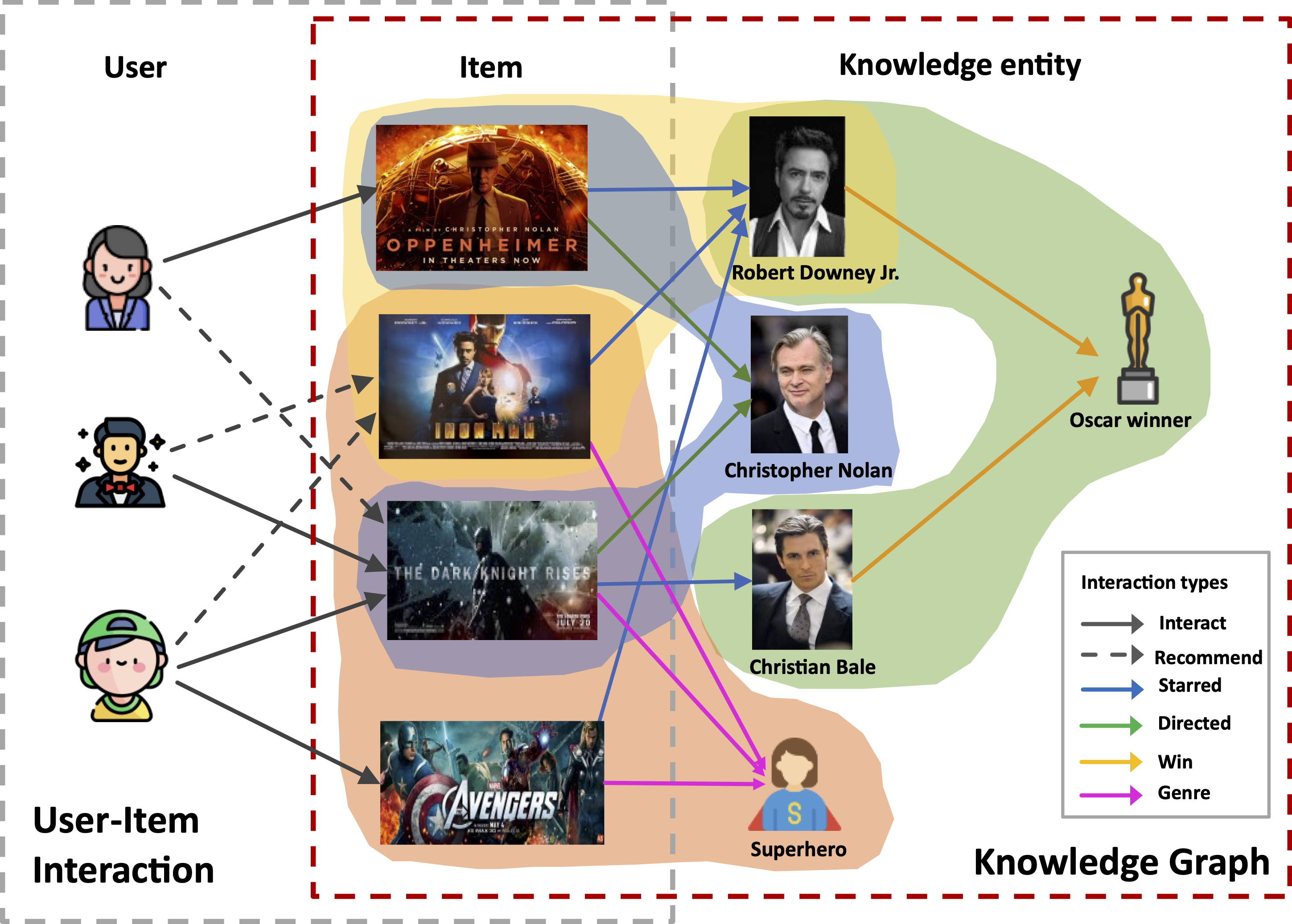}
	\caption{An illustration of knowledge-assisted recommender systems.}
	\label{fig:case_study}
\end{figure}

In this section, we present an example of knowledge-assisted recommendation, depicted in Fig.~\ref{fig:case_study}. For this example, we use data from the MovieLens dataset, a collection of movie ratings and interactions. This dataset is published by Grouplens~\footnote{http://www.grouplens.org/}. To add depth to our understanding, we connect the movies in this dataset to Freebase, a vast repository of structured information. Freebase aggregates detailed movie information, including the actors, directors, and genres, from reliable sources like Wikipedia and IMDB. 

In this example, three users exhibit interest in four films: ``\textit{Oppenheimer}", 
``\textit{Iron Man}, ``\textit{The Dark Knight Rises}", and ``\textit{Avengers}". Utilizing additional knowledge from the knowledge graph, we categorize the films into distinct groups, each represented by a color. Specifically, the "blue" group comprises films directed by \textit{Christopher Nolan}, the ``red" group includes movies within the \textit{Superhero} genre, the ``yellow" group features films starring \textit{Robert Downey Jr.}, and the "green" group encompasses movies having actors who are \textit{Oscar winners}. Unlike the traditional recommendation system, which only focuses on user interaction similarity, our knowledge-assisted system leverages this group information to derive high-order recommendations and suggest more meaningful choices.  For instance, if a user has interacted with  ``\textit{Oppenheimer}", starring Oscar winner \textit{Robert Downey Jr.} and directed by \textit{Christopher Nolan}, the system might recommend ``\textit{The Dark Knight Rises}", which shares the same director and starring \textit{Christian Bale}, who also share the Oscar-winning status. Similarly, a user interested in ``\textit{The Dark Knight Rises}" and ``\textit{Avengers}" may find the movie ``\textit{Iron Man}" suggested due to its shared \textit{Superhero} genre and leading actor, which is \textit{Robert Downey Jr.}

\subsection{Problem Formulation}
This section organizes key notations used throughout the paper and formalizes the problem of the recommender system with collaborative and knowledge signals.

\sstitle{User-item interaction}  To perform recommendations, in most cases, we exploit the information history of interaction between users and items such as the records of clicks, watches, and purchases. Such interaction data can be formed as user-item bipartite graph $\mathcal{G}_1$ which is compromised with a set of triplets $\{(u,y_{ui},i)|u \in \mathcal{U},i \in \mathcal{I})\}$ where the nodes are two disjoint sets $\mathcal{U}$ and $\mathcal{I}$(i.e, user and item sets). This can be represented in a matrix format $\mathcal{Y}$, where the number of rows and columns corresponds to the number of users and items respectively. Each entry of the interaction matrix $y_{ui}= 1$ if the user $u$ has interacted with item $i$, otherwise $0$.

\sstitle{Knowledge graph} As extra supportive information, we leverage knowledge graph(KG) which represents the intricate relationships between real-world objects based on its own atomic unit, namely triplet. Formally, each triplet follows the form of $\{(h,r, t)|h, t \in \mathcal{E},r \in \mathcal{R}\}$, where $\mathcal{E}$ and $\mathcal{R}$ indicate set of entities and relations respectively. For example, as it is shown in Figure.~\ref{fig:case_study}, a triplet (\textit{Peter Jackson, isDirectorOf, Lord of the Rings}) states the fact that Peter Jackson is the director of the movie \textit{Lord of the Rings}. It is worth noting that $\mathcal{R}$ involves both canonical(e.g., \textit{isDirectorOf}) and inverse direction(e.g., \textit{isDirectedBy}). Based on the richer connections between items and their attributes, the interpretability and quality of recommendations can be enhanced and more intelligible. For instance, a recommendation of a book can be generated based on its auxiliary information including but not limited to author, genre, or publisher.

\begin{myprob}[KG-enhanced recommendation] Given  the knowledge graph $\mathcal{G}_2$, set of users $\mathcal{U}$, set of items $\mathcal{I}$, and the user-item interaction matrix $\mathcal{Y}$, the problem is to learn a recommendation function $\mathcal{F}=\left(u, i \mid \mathcal{Y}, \mathcal{G}_2, \Omega\right)$ that predicts the non-interacted item $i$ $\left(i \in \mathcal{I}\right)$ that user $u$ $\left(u \in \mathcal{U}\right)$ would inclined to engage with, where $\Omega$ denotes the learnable parameters of the model.
\end{myprob}

\begin{table}[!h]
\centering
\small
\caption{Important notations}
\label{tab:deftable}
\scalebox{0.8}{
\begin{tabular}{cl}
\toprule
\textbf{Symbols}        & \textbf{Definition}                                            \\ \midrule
$\mathcal{G}_1$, $\mathcal{G}_2$, $\mathcal{G}$ & user-item bipartite graph, knowledge graph, collaborative knowledge graph \\      
$\mathcal{U}$, $\mathcal{I}$, $\mathcal{R}$, $\mathcal{E}$ & user set, item set, relation set, knowledge entity set \\ 
$\mathcal{R}^{\prime}=\mathcal{R} \cup\{\textit{Interact}\}$ & collaborative knowledge relation set\\
$\mathcal{Y}$, $\mathbf{A}$, $\mathcal{B}$ & user-item interaction matrix, hypergraph incidence matrix, attention matrix \\ 
$G$ & heterogeneous hypergraph\\
$G_H$ & collaborative knowledge hypergraph\\
$G_u$ & user hypergraph snapshot\\
$G_i$ & item hypergraph snapshot\\
$G_e$ & collaborative knowledge hypergraph snapshot\\
$T_V$, $T_E$ & number of node types, number of hyperedge types\\
$V$, $E$ & set of nodes, set of hyperedges\\
$|E|$, $|V|$ & number of hyperedges, number of nodes in a hypergraph \\
$V_H$, $V_u$, $V_i$, $V_e$ & heterogeneous nodes, user nodes, item nodes, knowledge entity nodes\\
$E_H$, $E_u$, $E_i$, $E_e$ & heterogeneous  hyperedge, user hyperedge, item hyperedge, collaborative knowledge hyperedge\\
$X$ & node feature matrix \\
$\textbf{H}$ & hidden feature matrix \\
$W$ & hyperedge weight matrix\\
$w(e)$ & hyperedge weight of hyperedge e\\
\bottomrule
\end{tabular}
}
\end{table}

\subsection{Approach Overview}
\label{ssec:overview}
\begin{figure}[!h]
	\centering
	\includegraphics[width=1.0\linewidth]{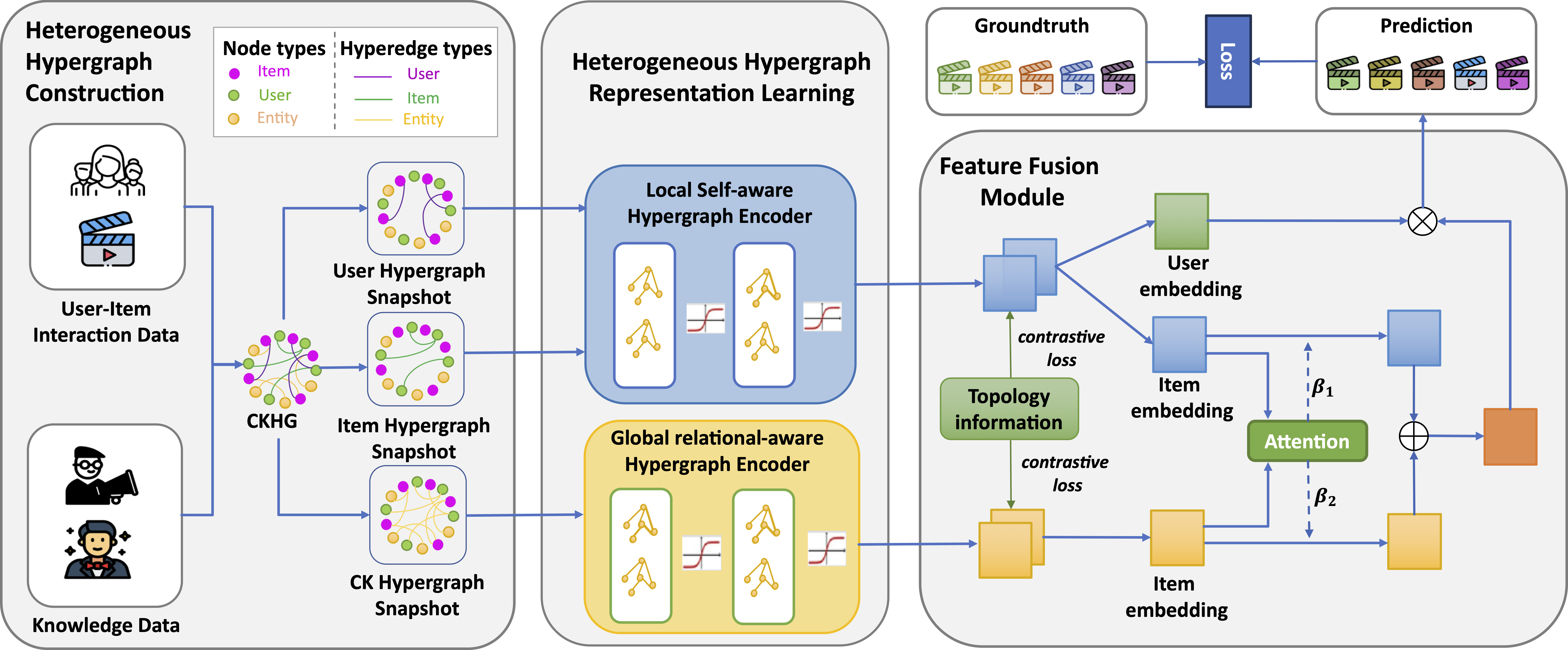}
	\caption{ Overview of our framework for KGHRec, which consists of three main components: Heterogeneous hypergraph construction, Heterogeneous hypergraph representation learning, and Feature fusion module.
}
	\label{fig:framework}
\end{figure}

In light of the above aspects, we propose a novel knowledge-assisted hypergraph recommendation system, as illustrated in Figure~\ref{fig:framework}. Overall, we propose a novel recommender system framework that unifies the learning of user-item interactions and knowledge signals. Initially, raw data, comprising user-item bipartite and knowledge graphs, are harnessed to formulate a heterogeneous hypergraph, serving to retain the group-wise characteristics inherent in the input network, thus addressing challenge \textbf{C1}. However, this data structure cannot model the complex relational dependencies between nodes in the input network. Consequently, we advocate for a relation-aware heterogeneous hypergraph attention encoder network to learn the hypergraph representations while capturing the attentive bias of relations towards instances, hence resolving \textbf{C2}. By joint learning the embedding of user-item interactions and knowledge graphs, our proposed method can exploit the high-order dependencies among users and knowledge entities. Moreover, we apply the relation-aware attention mechanism to generate the attentive score, which allows us to easily capture the reason why a specific item is chosen, hence solving \textbf{C3}. Finally, our proposed method uses different encoders to encode distinct aspects of the original network that include user, item, and knowledge entity nodes. The node embeddings retrieved from different encoders might include common information, such as users, items, or knowledge entities. Hence, we developed a two-step method, including attentive aggreation of embeddings and cross-view self-supervised training mechanism to solve \textbf{C4}.

To this end, we need to realize the following functions to instantiate the framework:\\
\sstitle{Heterogeneous hypergraph construction} As mentioned earlier, we combine the raw input data of the user-item bipartite graph and the knowledge graph to construct a heterogeneous hypergraph~\cite{tam2023wsdm}. Unlike the classical hypergraph, the heterogeneous hypergraph contains different types of nodes and multiple sizes of hyperedges so that it can alleviate the restrictive nature of the traditional hypergraph. Nodes can represent either users, items, or knowledge entities. Hence, we propose a method to construct the input for the hypergraph embedding process. 
Detailed insights into this component's construction are discussed in \autoref{sec:constructor}.

\sstitle{Heterogeneous hypergraph representation learning} The constructed heterogeneous hypergraph consists of several subgraphs, each representing a distinct hypergraph snapshot, namely user, item, and collaborative knowledge entity. 
The first two snapshots offer localized views of the user-item interaction graph, while the latter presents a global perspective, encompassing both user-item interactions and knowledge-related signals. 
To clearly distinguish between these two types of snapshots, we designate them as the ``\textit{Local}" and ``\textit{Global}" views of the heterogeneous hypergraph. The ``\textit{Local}" view focuses on micro-level examinations, specifically analyzing interactions and relationships between individual users and items. This allows for an in-depth exploration of specific elements within the hypergraph. On the other hand, the ``\textit{Global}" view adopts a more expansive scope. It extends beyond the immediate user-item interaction to include collaborative knowledge entities, integrating all subgraphs into a unified analysis.
To effectively learn from these ``\textit{Local}" and ``\textit{Global}" views, we introduce two tailored encoder networks: the \textit{Local self-aware hypergraph encoder} and the \textit{Global relational-aware hypergraph encoder}, respectively. The \textit{Local self-aware hypergraph encoder} integrates a self-attention mechanism with hypergraph embedding to capture the group-wise characteristics in user-item interactions and to highlight the significance of nodes within their neighborhood. Conversely, the \textit{Global relational-aware hypergraph encoder} merges a relational-aware attention mechanism with hypergraph embedding. This combination is designed to foster the understanding of relational effects among instances and to capture the high-order dependencies between users and knowledge entities. Comprehensive details on this learning module are provided in \autoref{sec:embed}.

\sstitle{Feature fusion module} To facilitate the learning of node embeddings from various hypergraph snapshots, we propose a two-step method that first employs an attention mechanism and then utilizes a cross-view self-supervised learning training scheme. This approach differs significantly from the attention mechanisms employed in graph encoders, which primarily focus on enhancing node embeddings. Specifically, the proposed attention mechanism functions as an aggregator and aims to combine latent features from different encoders. Meanwhile, the cross-view self-supervised learning scheme is designed to align similar entities' latent features across diverse latent spaces. We give more details about this component in \autoref{sec:contrastlearn}.

\section{Collaborative Knowledge Heterogeneous Hypergraph}
\label{sec:constructor}

This section first introduces the definition of related data structure, including collaborative knowledge graph (CKG), and heterogeneous hypergraph. After that, we introduce collaborative knowledge hypergraph (CKHG) - a novel data structure for unifying the user interaction and knowledge data and capturing the group-wise characteristics of the network, followed by a detailed description of how to construct it. 
The notation used is summarised in \autoref{tab:deftable}.

\subsection{Definition of collaborative knowledge heterogeneous hypergraph}
~\cite{wang2019kgat} first introduces the concept of CKG, a data structure designed to capture the high-order relationship between users and knowledge entities. Specifically, the detailed explanation of this data structure is illustrated in \autoref{def1}.

\begin{mydef}[Collaborative knowledge graph] 
\label{def1}
Let $\mathcal{G}$ represent the Collaborative Knowledge Graph (CKG), a structure encoding both user-item interaction signals and relational knowledge signals as a unified relational graph. The user-item interaction signals are derived from $\mathcal{G}_1=\{(u,y_{ui},i)|u \in \mathcal{U},i \in \mathcal{I})\}$, and the relational knowledge signals are obtained from $\mathcal{G}_2 =\{(h,r, t)|h, t \in \mathcal{E},r \in \mathcal{R}\}$. In CKG, each user behavior is meticulously represented as a triplet, ($u$, \textit{Interact}, $i$), wherein $y_{ui} = 1$  is symbolized as an additional relation, termed \textit{Interact}, existing between user $u$ and item $i$. Subsequently, the CKG is formally represented as $\mathcal{G}=\left\{(h, r, t) \mid h, t \in \mathcal{E}^{\prime}, r \in \mathcal{R}^{\prime} \right\}$, where $\mathcal{E}^{\prime}=\mathcal{E} \cup\mathcal{U}$ and $\mathcal{R}^{\prime}=\mathcal{R} \cup\{\textit{Interact}\}$.
\end{mydef}

While the CKG framework effectively integrates user-item bipartite and knowledge graphs into a unified graph, it overlooks the high-order interactions between users and knowledge entities. To address this, we propose using a data structure called heterogeneous hypergraph~\cite{sun2021heterogeneous}. This data structure, characterized by its diverse node and edge types, outperforms other data structures in terms of capturing complex group-wise characteristics. The specifics of this heterogeneous hypergraph are elaborated in \autoref{def2}.

\begin{mydef}[Heterogeneous Hypergraph]
\label{def2} Assume that $G(V, E)$ is a hypergraph that contains $|V|$ number of nodes and $|E|$ number of hyperedges. A positive edge weight $w(e)$ is given to each hyperedge $e \in E$. Note that the assigned weights are positioned in diagonal entries in a diagonal matrix $W \in R^{|E| \times |E|}$. Information about the association of nodes with hyperedges is formalized with incidence matrix $\mathbf{A} \in R^{|V| \times |E|}$ which is a binary rectangular matrix, where $\mathbf{A}_{ie}$ equals to 1 if node $v_i$ is part of the hyperedge $e$, otherwise 0. The sets $V$ and $E$ encompass ${T}_{V}$ and ${T}_{E}$ types of nodes and edges, respectively. $G$ is classified as a heterogeneous hypergraph if either ${T}_{V}$ or ${T}_{E}$ is greater than 1, signifying the existence of multiple types of nodes and edges within the hypergraph.
\end{mydef}

\noindent We then introduce a novel data structure, named Collaborative Knowledge Heterogeneous Hypergraph (CKHG), specifically tailored for the knowledge-based recommender system challenges. CKHG combines user interactions and item knowledge entities into a unified heterogeneous hypergraph framework. Mirroring the CKG structure, user behavior is represented as a triplet, $\left(u, Interact, i\right)$, with the addition of a distinct relation type $Interact$ signifying the linkage between user $u$ and item $i$. The definition of this proposed data structure is described as follows:

\begin{mydef}[CKHG] 
\label{def3}
Let $G_H(V_H, E_H)$ represent the CKHG, where the number of node types $T_V$ and edge types $T_E$ are set to three, respectively. Specifically, the node set $V_H$ is defined as $V_H = \{ V_u, V_i, V_e\}$, where $V_u$, $V_i$, and $V_e$ represent user, item, and knowledge entity nodes, respectively. Likewise, the edge set $E_H$ is articulated as $E_H = \{ E_i, E_u, E_e \}$, with $E_i$, $E_u$, and $E_e$ corresponding to the item, user, and collaborative knowledge entity hyperedges, respectively. A fact in CKHG is encapsulated by a tuple $\left(h, r, t, S_t \mid h, t \in V_H, r \in \mathcal{R}^{\prime}\right)$ with $\mathcal{R}^{\prime}=\mathcal{R} \cup\{\textit{Interact}\}$. $S_t$ is the set of supporting pairs $\{\left(v_{i}, r_{i}\right)\}_{i=1}^{|S_t|}$ with $v_i$ and $t$ are neighbours in the same hyperedge, and $r_i \in \mathcal{R}^{\prime}$ is the corresponding relation of the triplet $\left(h, r_i, v_i\right)$. 

\end{mydef}

\subsection{Constructing hypergraph snapshots}
\label{ssec:ckhgconstruction}
\begin{figure}[!h]
    \centering
    \includegraphics[width=0.7\linewidth]{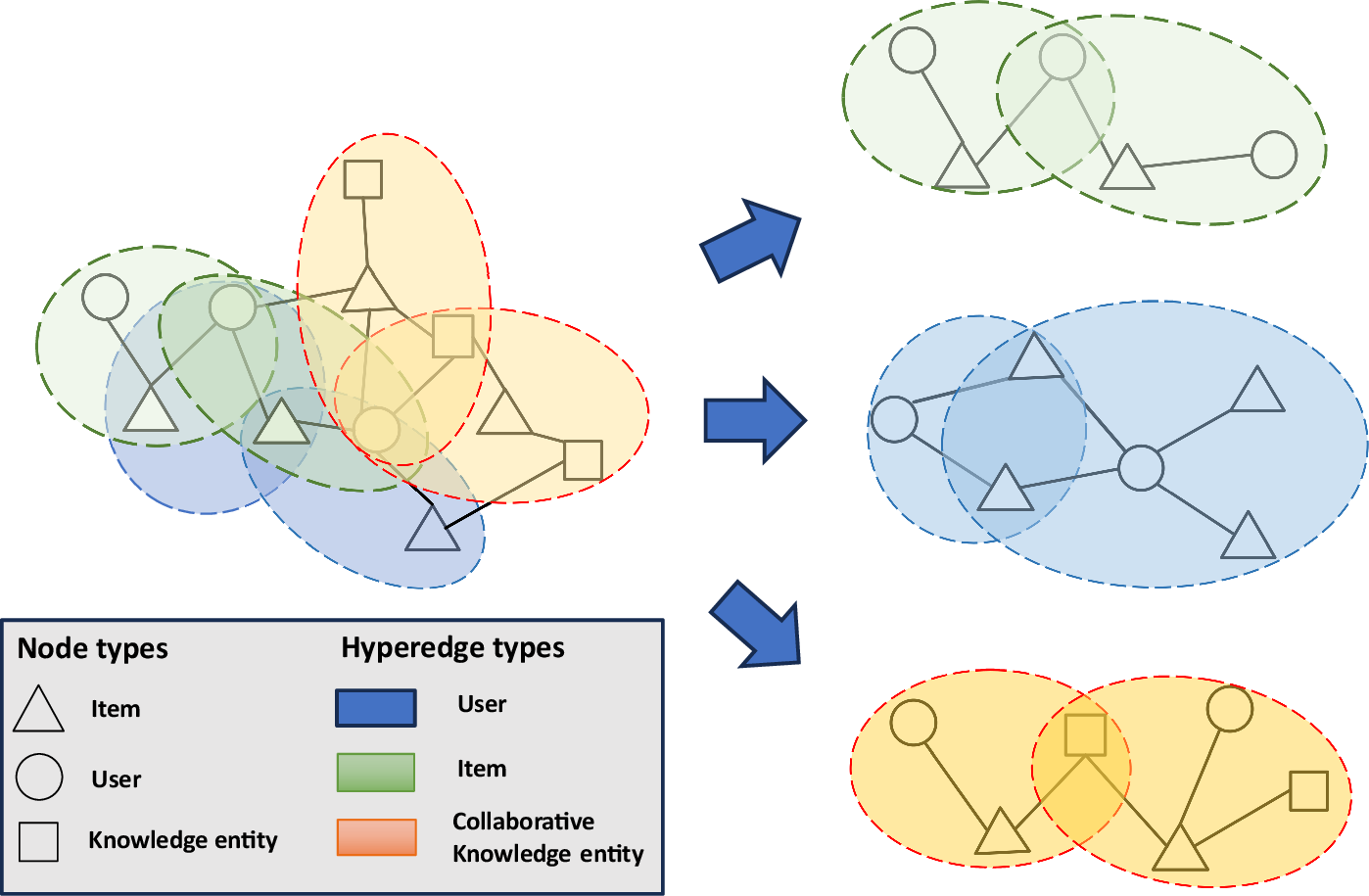}
    \caption{Snapshots generation for CKHG.}
    \label{fig:hypersnapshot}
\end{figure}

Given the input as a CKHG, we follow the approach from~\cite{sun2021heterogeneous} and decompose it into multiple hypergraph snapshots, with each snapshot encoding different information. Specificallly, the authors provide empirical evidence indicating that increasing the number of snapshots leads to a significant improvement in the model's convergence rate and the accuracy of its embeddings. This enhancement can be attributed to the ``\textit{divide and conquer}" training approach employed by the model. Under this framework, the model learns the embedding of each snapshot independently and concurrently, thereby reducing introduced noise and accelerating the learning process. Figure ~\ref{fig:hypersnapshot} illustrates the process of decomposing the original CKHG into multiple hypergraph snapshots.

\noindent Specifically, we define three types of hypergraph snapshots using the encoded information  as follows:
\begin{itemize}
    \item \textit{Item hypergraph:} Let $G_i(V_u, E_i)$ is a subgraph of $G_H$. Here $V_u \in V_H$ denotes the user nodes, and $E_i \in E_H$ denotes the item hyperedges, with a hyperedge $e_i \in E_i$ connecting users having interactions with item $i$.
    \item \textit{User hypergraph:} Let $G_u(V_i, E_u)$ is a subgraph of $G_H$. Here $V_i \in V_H$ denotes the item nodes, and $E_u \in E_H$ denotes the user hyperedges, with a hyperedge $e_u \in E_u$ connecting items having interactions with user $u$.
    \item \textit{Collaborative knowledge entity hypergraph:} Let $G_e(V_e, E_i \cup E_e)$ is a subgraph of $G_H$. This snapshot takes items and knowledge entities as hyperedges, with a hyperedge $e_k \in \left(E_i \cup E_e\right)$ connecting nodes $v_k \in V_H$ having relation $r \in \mathcal{R}^{\prime}$ with $e_k$.
\end{itemize}
By dividing the original big networks into smaller snapshots, we can learn these components simultaneously using different encoders, with each tailored for different purposes. This approach can lead to faster convergence of the model since our model now does not have to learn signals from different types of nodes and hyperedges, which often leads to suboptimal learned embedding. We then use different encoders to learn node embeddings in each snapshot and then aggregate these snapshots into a comprehensive representation. We divide the snapshots into two types based on the scope of the information encoded: \textit{Local view} - including user and item hypergraph snapshots, and \textit{Global view}, which is the collaborative knowledge entity hypergraph snapshot. While the former only encapsulates the user-item interaction, the latter captures both the user-item interaction and the item-entity relation while also demonstrating the high-order dependencies between the user and the knowledge entity. In order to learn the embedding of different types of snapshots, we use two different architectures: \textit{Local Self-aware Hypergraph Encoder} and \textit{Global Relational-aware Hypergraph Encoder}, which will be discussed in detail in \autoref{sec:embed}.

\section{Collaborative Knowledge Hypergraph Encoder}
\label{sec:embed}
This section depicts our methodology for encoding both collaborative signals and collaborative knowledge signals using two specially designed encoder networks: \textit{Local Self-aware Hypergraph Encoder} and \textit{Global Relational-aware Hypergraph Encoder}.

\subsection{Local Self-aware Hypergraph Encoder} 
In this section, we describe the architecture of the novel hypergraph convolution, e.g., \textit{Hypergraph Transformer Self-Attention Networks}, used in the proposed encoder, followed by the overall architecture of this architecture. We then proceed to give a detailed mathematical representation of the mentioned encoder.

\sstitle{Hypergraph Transformer Self-Attention Networks} Inspired by~\cite{unitransformer}, we propose the \textit{Hypergraph Transformer Self-Attention Networks} to capture the impact of neighboring nodes in the hypergraph effectively. We first apply the transformer architecture~\cite{transformer} to all nodes in the input hypergraph $\Tilde{G}$, where $\Tilde{G} \in \{G_u, G_i\}$ . Mathematically, the hypergraph $\Tilde{G}$ can be represented as:
\begin{equation}
    \Tilde{G} = (X, \mathbf{A}),
\end{equation}
where $X \in \mathbb{R}^{|\Tilde{V}| \times F}$ represents the node features matrix, $\mathbf{A} \in \mathbb{R}^{|\Tilde{V}| \times |E|}$ is the incidence matrix, with $|\Tilde{V}|$ is the number of nodes of the hypergraph $\Tilde{G}$.

\begin{figure}[!h]
	\centering
	\includegraphics[width=0.5 \linewidth]{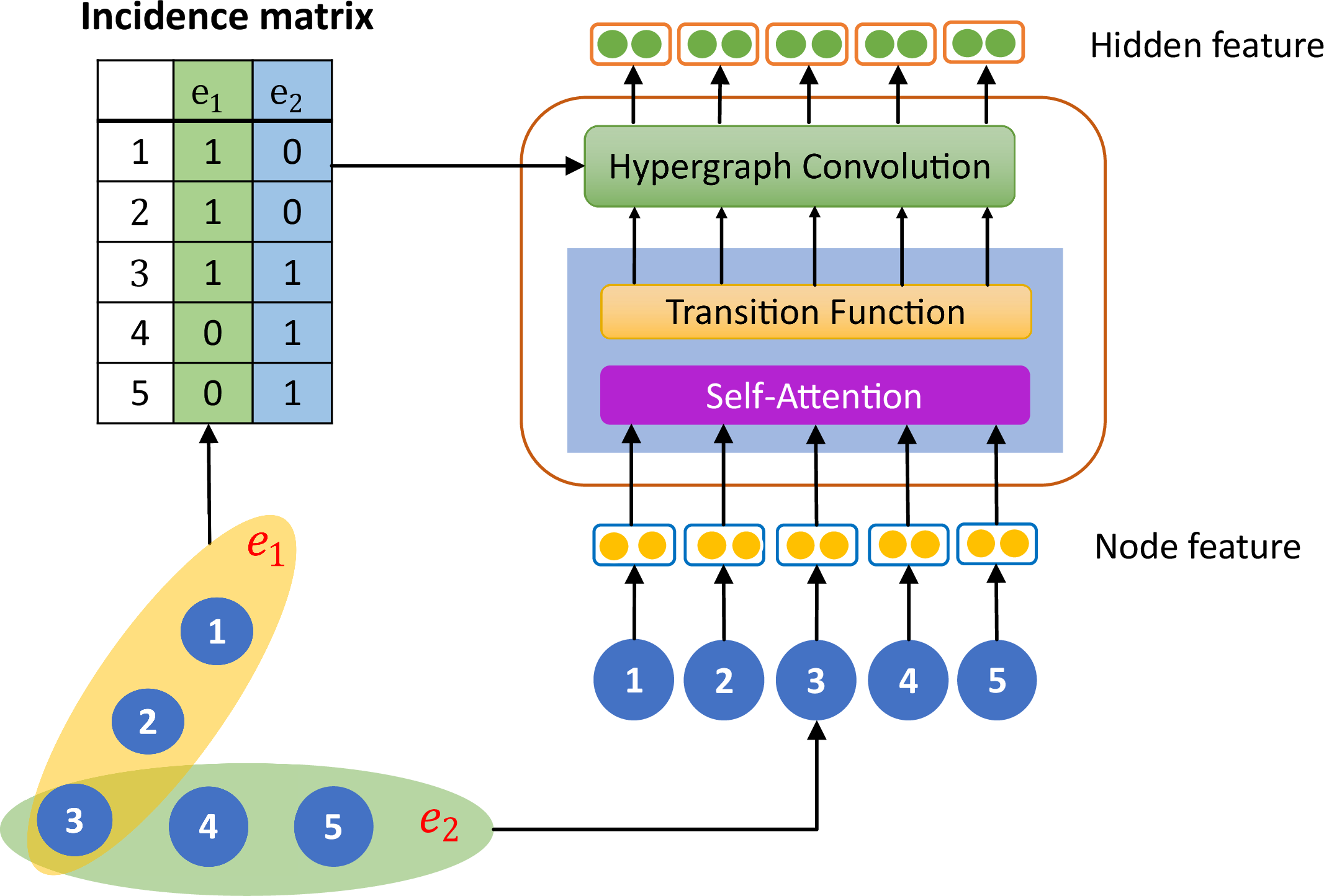}
	\caption{Architecture of Hypergraph Transformer Self-Attention Networks Convolution Layer.}
	\label{fig:hgtransformer}
\end{figure}

This design choice aims to leverage the self-attention mechanism to learn the node importance in the graph. Specifically, we update the node embeddings with their neighboring nodes by stacking multiple graph transformer layers with a weight-sharing mechanism. Each layer is composed of two key functions: the self-attention and the transition function. Nodes are first passed through self-attention function $\operatorname{ATT}(.)$ to weigh varied attention scores to different neighboring nodes, thus encoding dependencies between the current node $v$ and its neighbors. More precisely, the output of attention function $\operatorname{ATT-\mathbf{h}}_{t}^{(l)}$ at $l$-th layer and $t$ time step can be defined as below:
{\small
\begin{align}
  \label{eqn:transformer}
  \operatorname{ATT-\mathbf{h}}_{t,n}^{(l)} &= \text{Norm}\left(\mathbf{h}_{t-1,n}^{(l)}+\operatorname{ATT}\left(\mathbf{h}_{t-1,n}^{(l)}\right)\right),
\end{align}
}
{\small
\begin{align}
  \label{eqn:attention}
    \operatorname{ATT}\left(\mathbf{h}_{t-1,n}^{(l)}\right)=\sum_{n,\hat{n} \in \mathcal{N}_{v} \cup\{v\}} \alpha_{n,\hat{n}}^{(l)}\left(\mathbf{V}^{(l)} \mathbf{h}_{t-1,\hat{n}}^{(l)}\right)
\end{align}
}
{\small
\begin{align}
    \label{eqn:att}
    \alpha_{n,\hat{n}}^{(l)}=\operatorname{softmax}\left(\frac{\left(\mathbf{Q}^{(l)} \mathbf{h}_{t-1,n}^{(l)}\right)^{\top}\left(\mathbf{K}^{(l)} \mathbf{h}_{t-1, \hat{n}}^{(l)}\right)}{\sqrt{d}}\right),
\end{align}
}
where $n, \hat{n} \in \mathcal{N}_v \cup v$ are neighbor nodes of the given current node $v$, $\mathbf{h}_{n}^{(l)}$ denotes the vector representation of $n$ at $l$-th layer, $\text{Norm}$ indicates Layer Normalization, $\alpha_{n, \hat{n}}^{(l)}$ is attention score, $\mathbf{V}^{(l)}, \mathbf{Q}^{(l)}, \mathbf{K}^{(l)} \in \mathbb{R}^{d \times d}$ denotes linear projection matrices for value, query, and key respectively. 
The learned attention-aware intermediate parameters are then transferred to the Feed Forward Neural Network(FNN), which is denoted as Trans$(\cdot)$, followed by residual connection as below.
{\small
\begin{equation}
    \begin{split}
    \label{eqn:transition}
    \mathbf{Trans-h}_{t,u}^{(l)} &= \text{Norm}\left(\mathbf{ATT-h}_{t,n}^{(l)} 
+\operatorname{Trans}\left(\mathbf{ATT-h}_{t,n}^{(l)}\right)\right)
    \end{split}
\end{equation}
}
Note that, the topological information of input hypergraph $\Tilde{G}$ vanishes after the propagation through the self-attention layer since it naturally connects all possible pairs of nodes with all positions engaging in interactions with one another. To overcome this limitation, we propose using a hypergraph convolution HConv(·) after each transformer self-attention layer, as illustrated in Figure~\ref{fig:hgtransformer}.

Following~\cite{hgcn}, before the propagation through convolution layers, we need to construct Laplacian matrix $\mathbf{\Delta} \in \mathbb{R}^{d \times d}$ of hypergraph $\Tilde{G}$, which can be calculated as below:

\begin{equation}
\mathbf{\Delta} = \mathbf{I} - \mathbf{D_v}^{-1/2}\mathbf{A} \mathbf{D_e}^{-1}\mathbf{A}^T\mathbf{D_v}^{-1/2}
\end{equation}
where $\mathbf{A}$ denotes the incidence matrix, $\mathbf{D_v}$ and $\mathbf{D_e}$ are the degree matrices of nodes and hyperedges respectively.
In the remainder of this paper, we use term $\mathbf{\Theta}$ for $\mathbf{I} - \mathbf{D_v}^{-1/2}\mathbf{A} \mathbf{D_e}^{-1}\mathbf{A}^T\mathbf{D_v}^{-1/2}$ for clarity of presentation.
It is worth noting that the Laplacian matrix can further be transformed into diagonal matrix $\mathbf{\Delta}=\mathbf{\Psi}\mathbf{\Lambda}\mathbf{\Psi^T}$, where $\mathbf{\Psi}$ is eigenvector matrix and $\mathbf{\Lambda}$ is a diagonal matrix containing the eigenvalues of $\mathbf{\Delta}$ as its diagonal elements.
Hence, HGConv(.) is mathematically represented as:
\begin{align}
  \label{eqn:hgconv}
  \operatorname{HGConv}\left(X, \Theta, \mathbf{P}\right) = \sigma\left(\mathbf{\Theta} \cdot X \cdot \mathbf{P}\right),
\end{align}
where $\mathbf{P}$ is a learnable weight matrix. Formally, we define a Hypergraph Transformer Self-Attention Networks Convolutional (HGTNConv) layer as:
{\small
\begin{align}
  \label{eqn:transformer}
  \mathbf{H}^{\prime(l)}&=\operatorname{Transformer}\left(\mathbf{H}^{(l)} \mathbf{Q}^{(l)}, \mathbf{H}^{(l)} \mathbf{K}^{(l)}, \mathbf{H}^{(l)} \mathbf{V}^{(l)}\right),\\
  \mathbf{H}^{(l+1)}&=\operatorname{HGConv}\left( \mathbf{H}^{\prime(l)}, \mathbf{\Theta}, \mathbf{P}\right),
\end{align}
}
where $\textbf{H}^{(0)} = X$ is the node feature matrix.

We then stack multiple layers of HGTNConv on top of each other to construct the complete HGTN. The embedding operation of HGTN can be mathematically represented as follows:
{\small
\begin{equation}
    \begin{split}
    f(\Tilde{G}, \mathbf{A}) &= {\operatorname{HGTNConv}}^{(l)}  (...\operatorname{HGTNConv}^{(1)}(\textbf{H}^{(0)}, \mathbf{A}), \mathbf{A}), 
    \end{split}
\end{equation}
}%
where $l$ denotes the number of layers in HGTN.\\
Furthermore, to strengthen the output signal, we utilize the residual term after the HGTN layer following the design in~\cite{GraphBert}, which is denoted as $\operatorname{Res}$. The complete mathematical representation of the \textit{Local Self-aware Hypergraph Encoder} is as follows:
{\small
\begin{align}
    \mathcal{M}_i &= {\operatorname{HGTN}} (G_i) + {\operatorname{Res}}(X_i),\label{eq:localitemencoder}\\
    \mathcal{M}_u &= {\operatorname{HGTN}} (G_u) + {\operatorname{Res}}(X_u), 
    \label{eq:localuserencoder}
\end{align}
}%
where $G_i = \left(X_u, \mathbf{A}_i\right)$ represents the item hypergraph snapshot, $X_u \in \mathbb{R}^{|V_u| \times F}$ represents the user embedding, and $\mathbf{A}_i \in \mathbb{R}^{|V_i| \times |V_u|}$ represents the item incidence matrix, with $|V_i|, |V_u|$ represent the number of items and users, respectively. Analogously, $G_u = \left(X_i, \mathbf{A}_u\right)$ represents the user hypergraph snapshot, $X_i \in \mathbb{R}^{|V_i| \times F}$ represents the item embedding, and $\mathbf{A}_u \in \mathbb{R}^{|V_u| \times |V_i|}$ represents the user incidence matrix.

\subsection{Global Relational-aware Hypergraph Encoder} 

\begin{figure}[!h]
	\centering
	\includegraphics[width=0.8 \linewidth]{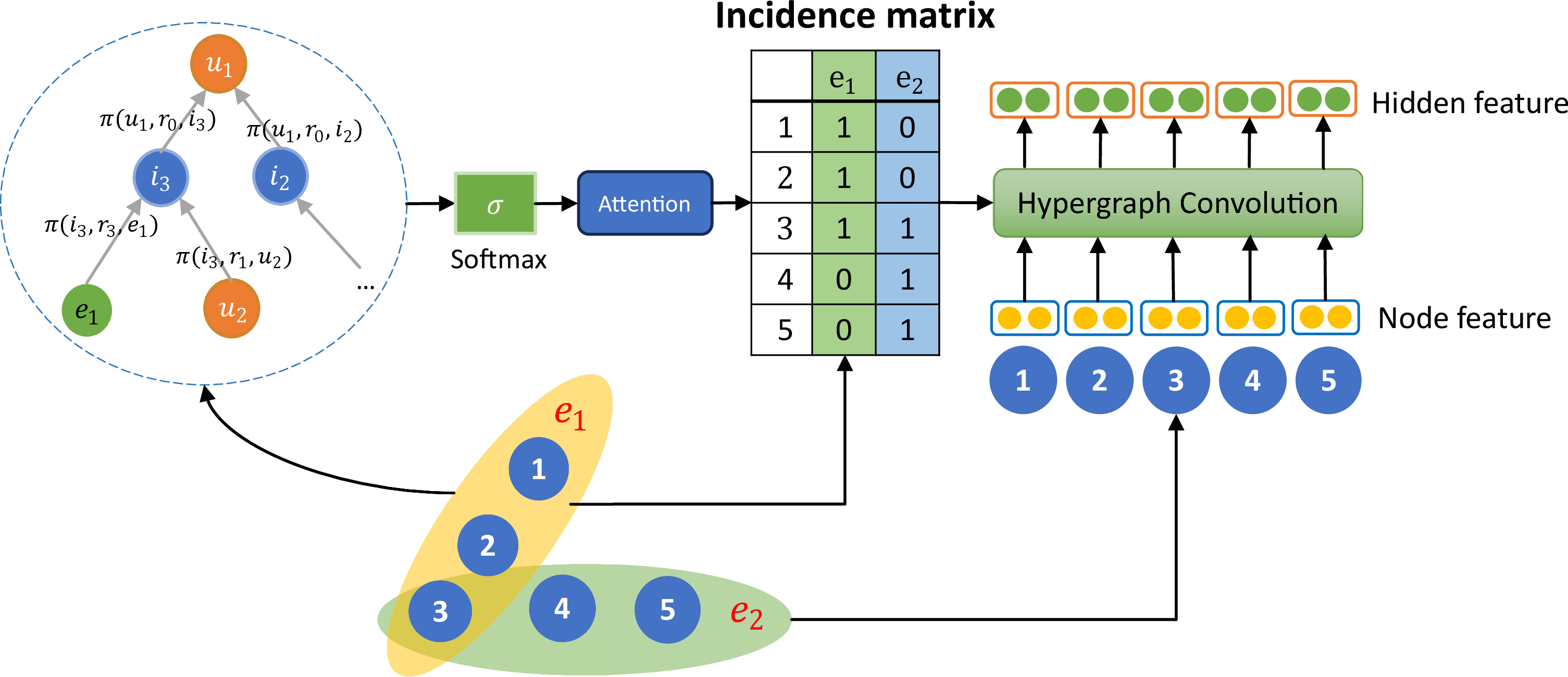}
	\caption{Architecture of Relational-aware Hypergraph Attention Convolution Layer.}
	\label{fig:relationalhgattention}
\end{figure}

Inspired by the graph attention mechanisms in~\cite{ kgat}, we design a relation-aware hypergraph attention layer to capture the relation heterogeneity over collaborative and knowledge graph connection structures (Figure \ref{fig:relationalhgattention}). Specifically, instead of using the cosine similarity to measure the impact of neighboring nodes as in~\cite{hgcn}, we tailor the attention matrix using the relational-aware attention mechanism, thus better expressing the relational dependency between users-items-entities.

\sstitle{Relational-aware Attention Mechanism} Given head entity $h$, the set of triples connected to $h$ forms ego-centric network, $\mathcal{N}_h = \{(h,r,t)|(h,r,t) \in G_e\}$, where $h$ is head node(i.e., ego node) and $t$ denotes the tail node~\cite{DeepInf}. 

Then, followed~\cite{wang2019kgat}, we denote the impact factor of tail entity $t$ regarding the relation $r$ to the head entity $h$ as $\Tilde{\pi}(h,r, t)$, with the mathematical formulation as follows:
{\small
\begin{equation}
    \Tilde{\pi}(h, r, t)=\left(\mathbf{W}_r \mathbf{e}_t\right)^{\top} \tanh \left(\left(\mathbf{W}_r \mathbf{e}_h+\mathbf{e}_r\right)\right),
\end{equation}
}
where $\operatorname{tanh}$ is the activation function, $\mathbf{W}_r$ is the learnable matrix. This results in the attention score being influenced by the proximity between $e_h$ and $e_t$ within the space of relation $r$, hence, allowing for more substantial information propagation for entities that are closer together.\\
Then, the softmax function is applied to normalize the learned attention scores:%
{\small
\begin{equation}
    \pi(h, r, t)=\frac{\exp (\Tilde{\pi}(h, r, t))}{\sum_{\left(h, r^{\prime}, t^{\prime}\right) \in \mathcal{N}_h} \exp \left(\Tilde{\pi}\left(h, r^{\prime}, t^{\prime}\right)\right)}.
\end{equation}
}

\sstitle{Relational-aware Hypergraph Attention Convolution} We then propose the \textit{Relational-aware Hypergraph Attention Convolution} (RHGATConv), where we employ the proposed attention mechanism to construct the attention matrix $\mathcal{B}$, which is mathematically represented as follows:
\begin{equation}
    \mathcal{B}_{ij} = \pi(h_i, r_{ij}, t_j),
\end{equation}
where $\mathcal{B}_{ij}$ represents the impact factor of the tail entity $t_j$ to the head entity $h_i$ regarding the relation $r_{ij}$, which acts as a gating mechanism to control how much information the tail entity can transfer its corresponding head entity. The attention matrix is then multiplied with the original hypergraph incidence matrix to create a more dynamic incidence matrix, thus better revealing the intrinsic relationship between vertices. The attention hypergraph incidence matrix is formally denoted as:
\begin{equation}
    \Tilde{\mathbf{A}} = \mathcal{B} \cdot \mathbf{A}.
\end{equation}
Denoting $\mathbf{D_v}^{\frac{1}{2}}\Tilde{\mathbf{A}}\mathbf{D_e}^{-1}\Tilde{\mathbf{A}}^T\mathbf{D_v}^{-1/2}$
 by $\Tilde{\mathbf{\Theta}}$, we formally define a RHGATConv layer as follows:
{\small
\begin{align}
  \label{eqn:rhgatconv}
  \mathbf{H}^{(l+1)}&=\sigma\left(\Tilde{\mathbf{\Theta}} \cdot\mathbf{H}^{(l)} \cdot \mathbf{P} \right),
\end{align}
}%
where $\textbf{H}^{(0)} = X$ is the feature matrix of all nodes in the input hypergraph, $\mathbf{A}$ is the incidence matrix, and $l$ denotes the $l$-th layer. We then stack multiple RHGATConv layers on each other to construct the complete RHGAT model. The embedding operation of RHGAT can be mathematically represented as follows:
{\small
\begin{equation}
    \begin{split}
        f(G_e, \mathcal{B}, \mathbf{A}) ={\operatorname{RHGATConv}}^{(l)}  (...\operatorname{RHGATConv}^{(1)}(\textbf{H}^{(0)}, \mathcal{B},\mathbf{A}), \mathcal{B},\mathbf{A}), 
    \end{split}
\end{equation}
}%
where $l$ denotes the number of layers in RHGAT.\\
We also utilize the residual term after the RHGAT layer. The complete mathematical representation of the \textit{Global Relational-aware Hypergraph Encoder} is as follows:
{\small
\begin{align}
    \mathcal{M}_{e} &= {\operatorname{RHGAT}} (X_{e}, \mathbf{A}_{e}) + {\operatorname{Res}}(X_{e}), \label{eq:globalencoder}
\end{align}
}%
where $\mathcal{M}_{e}$ denotes the collaborative knowledge latent feature, $G_e = \left(X_{e}, \mathbf{A}_{e}\right)$ represents the collaborative knowledge hypergraph snapshot, $X_{e} \in \mathbb{R}^{|V_e| \times F}$ represents the collaborative knowledge entity embedding; $\mathbf{A}_{e} \in \mathbb{R}^{|V_e| \times |V_e|}$ represents the incidence matrix of the corresponding hypergraph, $|V_e|$ represent the total number of collaborative knowledge entities, and $F$ represents the feature embedding size.

\sstitle{Semantic Representation Enhancement} To further encapsulate the relation-supported contextual information into the representation, we borrow the concept from $\operatorname{TransR}$~\cite{TransR} to learn the plausibility of each triplet in the graph. In particular, given a triplet $(h,r, t)$, the translation constraint $e_h^r + e_r \approx e_t^r$ enforces the head entity, which is translated with relation $r$ to be close as much as possible to the tail entity. Note that $e_h^r$ and $e_t^r$ indicate the entity embeddings of head and tail entities that are projected into the $r$ relation vector space. Formally, the energy of a triplet can be represented as follows:
\begin{equation}
    \delta(h, r, t)=\left\|\mathbf{W}_r e_h+ e_r-\mathbf{W}_r e_t\right\|_2^2,
\end{equation}
where $\mathbf{W}_r$ is the mapping matrix regarding the specific relation $r$.

The pairwise ranking objective is optimized to encourage the model to rank positive triplets higher than those that form a negative(i.e., invalid) relationship:
\begin{equation}
    \label{eqn:kg}
    \mathcal{L}_{\mathrm{KG}}=\sum_{\left(h, r, t\right) \in \mathcal{T} \left(h, r, t^{\prime}\right) \in \mathcal{T}^{\prime}}-\ln \sigma\left(d\left(h, r, t^{\prime}\right)-d(h, r, t)\right),
\end{equation}
where $\mathcal{T} = {(h,r, t)|(h,r, t) \in G_{e}}$ is a set of positive triplets, $\mathcal{T^{\prime}}$ represents the set of negative triplets that are constructed by corrupting the positive triplets from $\mathcal{T}$, $d$ denotes the distance function, and $\sigma$ denotes the nonlinear sigmoid activation function.

\section{Enhance Learned Embedding With Attention-aware Feature Fusion And Cross-view Contrastive Learning}
In this section, we elaborate on the design of the Attention-aware Feature Fusion Module and the Knowledge-guided Cross-view Contrastive Learning mechanism, followed by the loss function used in our proposed model.
\label{sec:contrastlearn}
\subsection{Attention-aware Feature Fusion Module}
The item signal can be retrieved from embedding $\mathcal{M}_i$ and $\mathcal{M}_e$. However, each embedding represents a different aspect of the item~\cite{dcah}. For example, $\mathcal{M}_i$ encodes the multi-order pairwise relationship of items with interacting users, while $\mathcal{M}_e$ encodes the multi-order relational dependencies among entities, including items. Hence, to effectively combine the item signal from these embeddings, we propose using an additional attention layer with the weight-sharing approach to acquire the optimal-weighted fusion of the sets of varied-purposed embeddings.\\
Specifically, we first retrieve the item latent feature from $\mathcal{M}_e$. For a clear explanation, we denote the item latent features retrieved from $\mathcal{M}_i$ as $\mathcal{M}_i^{cf}$ since it is retrieved from the collaborative filtering channel, and the item latent features retrieved from the collaborative knowledge hypergraph embedding $\mathcal{M}_e$ as $\mathcal{M}_i^{ckg}$. 
After that, we proceed to transform embeddings from both channels to a non-linear transformation such as a single-layer $\operatorname{MLP}$.
Following that, we assess the weight of each channel-specific embedding by gauging the similarity with a channel-level attention vector $\mathbf{q}$.
Then, for the final step, we compute the mean of the entire channel-wise item representations to get the final weight coefficient $w_{\phi_c}$ as follows:
{\small
\begin{equation}
    w_{\phi_c}=\frac{1}{|\mathcal{V}|} \sum_{i \in \mathcal{V}} \mathbf{q}^T \cdot \tanh \left(\mathbf{W} \cdot \mathcal{M}_i^{\phi_c}+\mathbf{b}\right), \phi_c \in\{cf, ckg\},
\end{equation}
}%
 where $\mathcal{V}$ denotes the set of items verticies, $\mathcal{M}_i^{\phi_c}$ denotes channel-specific embedding, $\mathbf{W}$ denotes the trainable weight matrix, and $\mathbf{b}$ denotes bias vector. The softmax function is applied to the weights of each channel $\phi_c$ to obtain a normalized attention score for each channel so that it can ensure the probability distributions over the contribution of each channel.
{\small
\begin{equation}
    \beta_{\phi_c}=\frac{\exp w_{\phi_c}}{\sum_{\phi_c} \exp w_{\phi_c}}, \phi_c \in\{cf, ckg\}
\end{equation}
}%
Here, $\beta_{\phi_c}$ represents the learned channel-wise attention weight.
Given the attention weights, we assign different importance to individual channel embeddings and then fuse them as follows:
{\small
\begin{equation}
\textbf{M}_i=\beta_{\phi_{cf}} \cdot \mathcal{M}^{cf}_{i}+\beta_{\phi_{ckg}} \cdot \mathcal{M}^{ckg}_{i}
\label{eq:attention}
\end{equation}
}%
\subsection{Knowledge-guided Cross-view Contrastive Learning}
\sstitle{Data augmentation on graph structure}  
Following~\cite{xia2022hypergraph}, we apply the DropEdge operation to create the augmented graph. The random selection of DropEdge enables the augmentation of the input graph by eliminating a specific number of edges iteratively during the training process, which resembles the self-supervised learning approach. 
This operation prevents the parameters from being overfitted to the training set and further addresses the challenge of over-smoothing~\cite{dropedge}. 
Specifically, the edge dropout process for the hypergraph structure is as follows:
\begin{equation}
    \mathcal{\mathbf{A}} \leftarrow \mathcal{Z}_{\mathbf{A}} \circ \mathbf{A},
\end{equation}
where $\leftarrow$ performs assignment operation and $\circ$ represents the element-wise multiplication, $\mathcal{Z}_{\mathbf{A}}$ denotes the binary mask matrix that drops out edges with a certain probability, and $\mathbf{A}$ indicates hypergraph incidence matrix.

\sstitle{Cross-view Contrastive learning} 
As aforementioned, we generate two separate embeddings that encapsulate different contexts: i) local collaborative relation encoding over the user and item hypergraph and ii) global relation-aware structure learning among collaborative knowledge hypergraph. Ideally, we want latent features of an item from different latent spaces to be in close proximity to each other in the embedding spaces.
Hence, we use a cross-view contrastive learning mechanism to reinforce this constraint.\\
Given augmented input graphs, we consider the latent features of a specific item from two vector spaces as a valid pair and that of a different item as an invalid pair. Based on this assumption, the proposed contrastive loss is minimized to encourage the latent vectors of an identical item to be close to each other; otherwise, they are located apart. Mathematically, our contrastive loss can be represented as below by following InfoNCE~\cite{InfoNCE}.

\begin{equation}
    \label{eqn:ssl_u}
    \mathcal{L}_s^{(u)}=\sum_{i=0}^I \sum_{l=0}^L-\log \frac{\exp \left(s\left(\mathbf{z}_{i, l}^{(u)}, \Gamma_{i, l}^{(u)}\right) / \tau\right)}{\sum_{i^{\prime}=0}^I \exp \left(s\left(\mathbf{z}_{i, l}^{(u)}, \Gamma_{i^{\prime}, l}^{(u)}\right) / \tau\right)}
\end{equation}

Here, $s(\cdot)$ denotes the cosine similarity function to measure the distance in the vector space, and $\tau$ indicates a hyper-parameter, namely the temperature parameter. $\mathbf{z}_{i, l}$ represents the hypergraph-guided collaborative latent vector of item $i$ in layer $l$, and $\Gamma_{i, l}$ denotes the latent vector of the identical item in the same layer from the global hypergraph-aware collaborative knowledge embedding.
Similarly, our contrastive loss for item representations is defined as:
\begin{equation}
    \label{eqn:ssl_i}
    \mathcal{L}_s^{(v)}=\sum_{i=0}^I \sum_{l=0}^L-\log \frac{\exp \left(s\left(\mathbf{z}_{i, l}^{(v)}, \Gamma_{i, l}^{(v)}\right) / \tau\right)}{\sum_{i^{\prime}=0}^I \exp \left(s\left(\mathbf{z}_{i, l}^{(v)}, \Gamma_{i^{\prime}, l}^{(v)}\right) / \tau\right)}
\end{equation}
This allows the local and global dependency views to supervise each other collaboratively, which enhances the user and item representations.\\
\sstitle{Model prediction} We obtain the representations for user, namely $\mathbf{e}_u$, from the local user embedding $\mathcal{M}_u$; while the item node $i$, namely $\mathbf{e}_i$ is obtained from the item embedding retrieved from the \textit{Attention-aware Feature Fusion Module} $\textbf{M}_i$. To measure the matching probability, we apply the inner product between all user and item representations as below,
\begin{equation}
    \label{eqn:predict}
    \hat{y}(u,i) =  \mathbf{e}_u^{\mathbf{T}} \mathbf{e}_i,\\
\end{equation}

where $\mathbf{e}_u \in \mathcal{M}_u$ and $\mathbf{e}_i \in \textbf{M}_i$ indicate user vector $u$ and item vector $i$, respectively. Additionally, we optimize the CF model by minimizing the score of $\operatorname{BPR}$ loss~\cite{BPR} function. The main concept behind the loss function is to assign a higher prediction score to the observed user preference than those of unobserved interactions.
\begin{equation}
    \label{eqn:bpr}
    \mathcal{L}_{\mathrm{CF}}=\sum_{(u, i) \in O (u, j) \in O^{\prime}}-\ln \sigma(\hat{y}(u, i)-\hat{y}(u, j)),
\end{equation}
where $(u, i) \in \mathcal O$ denotes the set of observed interaction of user $u$, $(u, j) \in  O^{\prime}$ represents the set of corrupted interaction of $u$ that has not been observed and $\sigma(\cdot)$ denotes the sigmoid activation function.

Finally, we have the objective function to learn Equations~\ref{eqn:bpr}, \ref{eqn:kg}, \ref{eqn:ssl_u}, and \ref{eqn:ssl_i} jointly, as follows:
\begin{equation}
    \label{eqn:loss}
    \mathcal{L} = \mathcal{L}_{CF} + \mathcal{L}_{KG} +  \lambda_2 \cdot (\mathcal{L}_s^{(u)} + \mathcal{L}_s^{(v)}) + \lambda_1 \cdot \left\|\Omega \right|^2_{\mathrm{F}}
\end{equation}

The overall training process of our framework is summarized in the Algorithm \ref{alg:training_process}.

\begin{algorithm}[!h]
\small
\SetKwInput{KwInput}{Input}
\SetKwInput{KwOutput}{Output}
\SetKwRepeat{KwRepeat}{Repeat}
\DontPrintSemicolon

\KwInput{User-item interaction matrix $\mathcal{Y}$, knowledge graph $\mathcal{G}_2$, set of users $\mathcal{U}$, set of items $\mathcal{I}$}
\KwOutput{Recommendation function $\mathcal{F}=\left(u, i \mid \mathcal{Y}, \mathcal{G}_2, \Omega\right)$, where $i \in \mathcal{I}$ and $u \in \mathcal{U}$}
  
Construct CKHG $G_H = \left( V_H, E_H\right)$ as in  \autoref{ssec:ckhgconstruction};

\Repeat{convergence}{
    Compute local hypergraph embeddings $\mathcal{M}_i$ and $\mathcal{M}_u$ using Equations \ref{eq:localitemencoder} and \ref{eq:localuserencoder};
    
    Compute the knowledge entity hypergraph embedding $\mathcal{M}_e$ using Equation \ref{eq:globalencoder};
    
    Retrieve $\mathcal{M}_i^{ckg}$ from $\mathcal{M}_e$;
    
    Calculate final item embedding $\mathbf{M}_i$ using Equation \ref{eq:attention};
    
    Compute contrastive losses $\mathcal{L}_{s}^{(u)}$ and $\mathcal{L}_{s}^{(v)}$ using Equations \ref{eqn:ssl_u} and \ref{eqn:ssl_i};
    
    Determine model loss using Equation \ref{eqn:loss} and update model parameters;
}
\caption{Training Process}
\label{alg:training_process}
\end{algorithm}

\section{Empirical Evaluation}
\label{sec:exp}

In this section, we empirically evaluate our framework based on the following research questions:

\begin{compactitem}
\item[\textbf{RQ1:}] Does our model outperform the baseline methods?
\item[\textbf{RQ2:}] How is the running time of the proposed model compared to other baselines? 
\item[\textbf{RQ3:}] How well can our model perform in the cold-start setting?
\item[\textbf{RQ4:}] How well can our model cope with noise in training data? 
\item[\textbf{RQ5:}] Does our model work well in limited data settings?
\item[\textbf{RQ6:}] What is the influence of each model component?
\item[\textbf{RQ7:}] How is the model interpretation ability of our KHGRec?
\item[\textbf{RQ8:}] Is our model sensitive to hyperparameters?
\end{compactitem}

In this section, we first describe the experimental setting (\autoref{ssec:exp_setup}). We then present our empirical evaluations of recommendation tasks on different datasets, including an end-to-end comparison (\autoref{sec:end2end}), a runtime analysis(\autoref{ssec:runtime}), a cold-start analysis(\autoref{ssec:coldstart}), a noise-resilient capability test(\autoref{ssec:noiseresilient}), a training data efficiency study(\autoref{ssec:trainingproportion}),  an ablation test (\autoref{ssec:exp_ablation}), a qualitative study (\autoref{ssec:exp_qualitative}), and an examination of the hyperparameter sensitivity (\autoref{ssec:exp_sensitivity}).

\subsection{Experimental Setting}
\label{ssec:exp_setup}

\sstitle{Datasets} 
To ensure a comprehensive and inclusive assessment, we undertake evaluations using four distinct real-world datasets: LastFM for music recommendations, MovieLens for film recommendations, Mind-F for news recommendations, and Alibaba-Fashion for shopping recommendations. By considering the varying degrees of sparsity in real-world interactions and the diversity of item knowledge within these datasets, we aim to illustrate the robustness of our model across a range of recommendation dataset accessibility scenarios. The comprehensive statistics of the datasets are summarized in \autoref{tab:datasets1}. 

\begin{table}[!h]
\centering
 \small
\caption{Statistics of Recommendation Datasets}
\label{tab:datasets1}
\begin{tabular}{c c c c c c}
\toprule
Statistics & LastFM & MovieLens-1M& Mind-F & Alibaba-Fashion\\ 
\hline 
\#Users & 1,892 & 2,000 & 100,000 & 114,737\\ 
\#Items  & 17,632 & 3,543 & 30,577 & 30,040\\ 
\#Interactions & 92,834 & 331,792 & 2,975,319 & 1,781,093 \\ 
\#Density & $2.7e^{-3}$ & $4.6e^{-2}$  & $9.7e^{-4}$ & $5.2e^{-4}$\\ 
\midrule
Knowledge Graph\\
\midrule
\#Entities & 399,910 & 49,774 & 24,733 & 59,156 \\ 
\#Relations & 18 & 49 & 512 & 51 \\ 
\#Triplets & 671,233 & 385,923 & 148,568 & 279,155 \\ 
\bottomrule
\end{tabular}
\end{table}

\begin{itemize}
    \item \textbf{LastFM} is a music recommendation dataset that extracts user and artist interaction information from the LastFM online music system and contains 1892 user ids and 17632 music artists with 92834 explicit interactions between them. The knowledge graph is established by connecting the items to the corresponding entities in Freebase. The ratio of actual interaction among all possible interactions is around $2.7e^{-3}$, which indicates that the number of observed interactions is sparse. 
    
    \item \textbf{Movielens} contains 2000 users and 3543 numbers of movie items with around 330 thousand explicit interactions with ratings ranging between 1 to 5. The density of the dataset demonstrates the deficiency of the amount of supervision information in the dataset. We then construct the knowledge graph by connecting the item with the corresponding entities in Freebase, such as director, soundtrack, nomination, etc. MovieLens-1M is published by Grouplens~\footnote{http://www.grouplens.org/}. 
    
    \item \textbf{Mind-F} consists of interactions between 100,000 users and 114,737 items that reflect users' preference behavior on news articles. The number of observed interactions over the number of all possible interactions is large compared to the other datasets adopted for the experiment by reaching $9.7e^{-4}$ density. The knowledge graph is constructed by extracting knowledge triplets from Wikidata based on the method used in \cite{tian2021joint}.

    \item \textbf{Alibaba-Fashion} is a dataset for fashion recommendation that contains interactions between 30,040 fashion outfit items and 114,737 users which originate from Alibaba's e-commerce platform. Each outfit item consists of various fashion pieces, each labeled according to a specified category. The number of interactions is $2,975,319$, and the density of the dataset is $5.2e^{-4}$, which signifies the sparsity of the dataset. We manually build the knowledge graph by constructing triplets that connect the item to its category following the approach leveraged in \cite{wang2021learning}.
    
\end{itemize}

\sstitle{Baselines} 
We select nine numbers of SOTA baselines to compare their performance with our model.
The baselines mainly cover three main research paradigms of current literature: GCNs-based, hypergraph-based, and KG-enhanced. Each category of baselines covers different facets of our model. Hence, the comparison between the chosen baselines and our proposed model can highlight the significance of our model in handling varied challenges to be solved in each research paradigm. By including baseline models from these three distinct categories, we aim to comprehensively evaluate our proposed model's performance across different promising recommendation paradigms.

\begin{itemize}
    \item \textbf{LightGCN}\cite{he2020lightgcn} introduces the simplified version of the GCN framework for CF by subducting redundant components that lie in the inherent design of existing GCN algorithms to make the model more suitable and robust to the recommendation tasks.

    \item \textbf{SGL}\cite{sgl} proposes self-supervised learning for CF tasks, which enriches the node representation by exploiting three different topological features as self-supervision signals.
    
    \item \textbf{DHCF}\cite{ji2020dual} proposes dual-channel hypergraph embedding propagation to retain the features of users and items individually while interconnecting them through the message-passing mechanism.

    \item \textbf{HCCF}\cite{xia2022hypergraph} aims to encode shallow and large scope of connectivity context into embeddings with GCN and hypergraph message-passing strategies. To address the scarcity problem of supervision data, the pipeline is trained with a contrastive learning schema.
    
    \item \textbf{SHT}\cite{sht2022} attempts to model global collaborative signals along with the user-item interactions using transformer architecture with a hypergraph attention scheme. The model augments the data through a self-augmented learning method that enforces local and global-aware embeddings to complement each other.
    
    \item \textbf{KGAT}\cite{wang2019kgat} applies the relation-aware attention mechanism on a KG-enhanced collaborative graph to distinguish the relevance of entities. The embeddings of nodes are propagated through the layers with their assigned attentive weights.

    \item \textbf{KGIN}\cite{wang2021learning} explores the user intent behind the collective user behaviors and incorporates the relation sequences for interaction and relation triples in KG separately to keep their own properties.

    \item \textbf{KGCL}\cite{kgcl} mitigates the noise in KG through the knowledge augmentation module and contrastive learning while learning the characteristics of user-item interaction signals.  
    
    \item \textbf{KGRec}\cite{kgrec} devises the generative learning model to capture the rationality of triplets in KG by training the model to reconstruct masked triplets that are weighted by a high rationality score. The model further alleviates the latent noisy knowledge based on the rationality score to prevent potential accuracy inferior.
    
\end{itemize}

\sstitle{Metrics} 
We assess the models based on how well they predict the user interest score on the candidate items by verifying the quality of top-k recommended items. To this end, we select two broadly-adopted ranking metrics: Recall@k and NDCG@k.

\begin{itemize}
    \item \textbf{Recall@k(Recall at k):} This metric measures the number of relevant items that appear in the list of top-k recommended items as below.
        \begin{equation}
            Recall@k = \frac{I_r \cap R_k}{I_r},
        \end{equation}
    where $I_r$ denotes the total number of relevant items in the dataset, and $R_k$ depicts the set of the top-k recommended items.
    
    \item \textbf{NDCG@k(Normalized Discounted Cumulative Gain at k):} To verify the quality of the ordering of the ranked recommendations, we utilize NDCG@k, which takes the position of recommendations in the list and the relevance of items at the same time. NDCG@k can be calculated by following the equation below.
    \begin{equation}
        NDCG_{@K} = \frac{DCG_{@K}}{IDCG_{@K}},
    \end{equation}
    where $DCG_{@K}$ indicates discounted cumulative gain, which assigns weights to the recommendation score based on the ranking position of recommended items, and $IDCG_{@K}$ denotes the ideal DCG score.
\end{itemize}
In the experimental settings, we specifically set the $k$ value to 10, 20, and 40.

\sstitle{Hyperparameter tuning}
Regarding the hyperparameters, we utilize the same settings from the original papers. The models are all optimized using the Adam optimizer. Concerning the hyperparameter of the proposed model, the learning rate is adopted from $\{1e^{-1}, 1e^{-2}, 1e^{-3}\}$. During the training, we exploited the learning rate scheduler to adjust the learning rate if no improvement was observed within 20 epochs for fast convergence. Batch sizes are set to 4096 for user-item interaction and  8192 for the KG dataset, respectively. We set the maximum number of epochs as 500 for the proposed and benchmark models. Temperature parameter $\tau$ is chosen from $\{0.1, 0.2, 1, 2, 10\}$. The $L_2$ regularization term is chosen from $\{1, 0.1, 1e^{-2}, 1e^{-3}, 1e^{-4}\}$. The contrastive regularization term is selected from $
\{0.1, 1e^{-2}, 1e^{-3}, 1e^{-4}, 1e^{-5}\}$. Finally, we also vary the number of layers from $\{1, 2, 3, 4\}$. We then further analyze the sensitivity of the hyperparameters in \autoref{ssec:exp_sensitivity}.

\sstitle{Reproducibility environment} 
We conduct experiments on each model 5 times with training, validation, and test sets that are split by the 7:1:2 ratio. The final performance results are averaged over the 5 runs to avoid any effect of randomness. The AMD Ryzen Threadripper 2950X system with 128GB RAM has been used for our evaluation. Our experiments are implemented in Pytorch, and codes can be found on our git repository~\footnote{https://github.com/vuviethung1998/KHGRec}.

\subsection{End-to-end comparison (RQ1)}
\label{sec:end2end}

In this section, we organize the overall performance evaluation results of KHGRec and competing methods. The summarization of the average performance over 5 times the runnings of the models is presented in \autoref{tab:baseline_evaluation}. 

\begin{table*}[!h]
\fontsize{25}{8}\selectfont
\centering
\caption{\textbf{The overall performance evaluation results for KHGRec and compared baseline models on two experimented datasets, LastFM and MovieLens-1M, where the best and second-best performances are denoted in bold and borderline, respectively.}}
\label{tab:baseline_evaluation}
\begin{adjustbox}{max width=1.02\linewidth}
\begin{tabular}
{@{}ccccccc@{\hskip -0.005in}ccccccc@{\hskip -0.01in}ccc@{}}\midrule \vspace{8pt}
\multirow{2}{*}{Model}  & \multicolumn{6}{c}{LastFM}&\phantom{}&\multicolumn{6}{c}{MovieLens-1M}\vspace{8pt}\\
\cmidrule{2-7} \cmidrule{9-14}
&Recall@10&NDCG@10&Recall@20&NDCG@20&Recall@40&NDCG@40&&Recall@10&NDCG@10&Recall@20&NDCG@20&Recall@40&NDCG@40\vspace{8pt}\\ \midrule
\text{LightGCN} & 0.1369 & 0.2228& 0.2007 & 0.2315 & 0.2814 & 0.2727 && 0.0968 & 0.355 & 0.1441 & 0.3064 & 0.2288 & 0.3033\vspace{8pt}\\
\text{SGL} & 0.1423 & 0.2327 & 0.2026 & 0.2345 & 0.2872 & 0.2778 && 0.0838 & 0.3093 & 0.1350 & 0.2928 & 0.2144 & 0.2896\vspace{8pt}\\
\midrule
\text{DHCF} & 0.1441 & 0.2265 & 0.2097 & 0.2423 & 0.2933 & 0.22847 && 0.0947 & 0.344 & 0.1557 & 0.3112 & 0.2502 & 0.3145\vspace{8pt}\\
\text{HCCF} & 0.1334 & 0.2126 & 0.2157 & 0.2406 & 0.3048 & 0.2852 && 0.1123 & 0.369 & 0.1807 & 0.3543 & 0.2874 & 0.3589\vspace{8pt}\\
\text{SHT} & 0.1614 & 0.2697 & 0.2384 & \underline{0.2624} & 0.3296 & \underline{0.3088} && 0.1332 & 0.3818 & 0.1836 & 0.3491 & 0.2856 & 0.3526\vspace{8pt}\\
\midrule
\text{KGAT} & 0.1582 & 0.2624 & \underline{0.2412} & 0.2589 & 0.3108 & 0.2775 && 0.1138 & 0.3393& 0.2112 & 0.3512 & 0.2782 & 0.333\vspace{8pt}\\
\text{KGIN} & 0.1685 & 0.2402 & 0.2369 & 0.2419 & 0.3333 & 0.2881 && 0.135& 0.3826 & 0.2169 & 0.3558 & 0.2965& 0.3542\vspace{8pt}\\
\text{KGCL} & 0.1652& 0.2423 & 0.2376 & 0.2472 & 0.324 & 0.2888 && 0.1343& \underline{0.3845} & 0.2143& 0.3645 & 0.3098& 0.3629\vspace{8pt}\\
\text{KGRec} & \underline{0.1686}& \underline{0.2732}& 0.2410 & 0.2493 & \underline{0.3355} & 0.2949 && \underline{0.1363}& 0.3832 & \underline{0.2171}& \underline{0.3692} & \underline{0.3106} & \underline{0.3747}\vspace{8pt}\\
\midrule
\textbf{KHGRec} &\textbf{0.1759}& \textbf{0.2764} & \textbf{0.2574} & \textbf{0.2809} & \textbf{0.3551} & \textbf{0.3302} && \textbf{0.1413}&\textbf{0.3965}&\textbf{0.2268}&\textbf{0.3958} & \textbf{0.3289} & \textbf{0.3968}\vspace{8pt}\\
\hline
\midrule
\textbf{\%Improve.} & 4.34\%& 1.17\% & 6.71\%& 7.05\%& 5.84\% & 6.93\% && 3.66\%& 3.12\%& 4.46\% & 7.2\% & 5.89\%& 5.9\%\\
\bottomrule
\end{tabular}
\end{adjustbox}
\end{table*}

\begin{table*}[!h]
\fontsize{25}{8}\selectfont
\centering
\caption{\textbf{The overall performance evaluation results for KHGRec and compared baseline models on two experimented datasets, Mind-F and Alibaba-Fashion, where the best and second-best performances are denoted in bold and borderline, respectively.}}
\label{tab:baseline_evaluation2}
\begin{adjustbox}{max width=1.02\linewidth}
\begin{tabular}
{@{}ccccccc@{\hskip -0.005in}ccccccc@{\hskip -0.01in}ccc@{}}\midrule
\multirow{2}{*}{Model}  & \multicolumn{6}{c}{Mind-F}&\phantom{}&\multicolumn{6}{c}{Alibaba-Fashion}\vspace{8pt}\\ 
\cmidrule{2-7} \cmidrule{9-14}
&Recall@10&NDCG@10&Recall@20&NDCG@20&Recall@40&NDCG@40&&Recall@10&NDCG@10&Recall@20&NDCG@20&Recall@40&NDCG@40\vspace{8pt}\\ \midrule
\text{LightGCN}  & 0.0165 & 0.0132 & 0.033 & 0.0222 & 0.0513 & 0.0291 && 0.0619 & 0.0481 & 0.0968 & 0.0609 & 0.1462 & 0.0657\vspace{8pt}\\
\text{SGL}  & 0.0152 & 0.0124 & 0.0276 & 0.0182 & 0.0459 & 0.0248 && 0.0624 & 0.0468 & 0.0856 & 0.0603 & 0.1476 & 0.0692\vspace{8pt}\\
\midrule 
\text{DHCF} & 0.0132 & 0.0133 & 0.0203 & 0.0154 & 0.0353 & 0.0206 && 0.0582 & 0.041 & 0.0762 & 0.0502 & 0.1328 & 0.0702\vspace{8pt}\\
\text{HCCF} & 0.0188 & 0.0146 & \underline{0.0369} & \underline{0.0227} & \underline{0.0602} & 0.0309 && 0.0625 & 0.0458 & 0.0987 & 0.0612 & 0.1505 & 0.0712\vspace{8pt}\\
\text{SHT} & 0.0174 & 0.014 & 0.0337 & 0.0201 & 0.0582 & \underline{0.0318} && 0.0635 & 0.0487 & 0.0942 & 0.0602 & 0.1487 & 0.0678\vspace{8pt}\\
\midrule 
\text{KGAT} & 0.0121 & 0.0097 & 0.0221 & 0.0158 & 0.0321 & 0.0215 && 0.0562 & 0.0382 & 0.0792 & 0.0552 & 0.1129 & 0.0657\vspace{8pt}\\
\text{KGIN} & 0.0158 & 0.0132 & 0.0277 & 0.0176 & 0.0473 & 0.0242 && \underline{0.066} & \underline{0.0495} & \underline{0.1024} & \underline{0.0626} & \underline{0.1536} & \underline{0.0777}\vspace{8pt}\\
\text{KGCL} & 0.0189 & 0.0138 & 0.0344 & 0.0213 & 0.0563 & 0.02867 && 0.0603 & 0.0472 & 0.0898 & 0.0579 & 0.1333 & 0.0707\vspace{8pt}\\
\text{KGRec} & \underline{0.0191} & \underline{0.0141} & 0.0362 & 0.0223 & 0.0534 & 0.0298 && 0.0627 & 0.0474 & 0.0974 & 0.0599 & 0.1463 & 0.0743\vspace{8pt}\\
\midrule
\textbf{KHGRec} & \textbf{0.0198} & \textbf{0.01576} & \textbf{0.0399} & \textbf{0.0246} & \textbf{0.0641} & \textbf{0.0334} && \textbf{0.0671} & \textbf{0.0503} & \textbf{0.1054} & \textbf{0.064} & \textbf{0.1575} & \textbf{0.0802}\vspace{8pt}\\
\hline
\midrule
\textbf{\%Improve.} & 3.4\%& 11.6\% & 10.8\%& 8.3\%& 6.3\% & 4.9\% && 1.44\%& 1.45\%& 2.88\% & 2.87\% & 2.5\%& 3.23\%\vspace{8pt}\\
\bottomrule
\end{tabular}
\end{adjustbox}
\end{table*}

\begin{itemize}
    \item \textbf{Overall Comparison:}  The results clearly demonstrate the superior performance of our model, which consistently outperforms the baseline methods in all tested scenarios across the entire datasets. This result highlights the efficiency of our proposed model compared to other baselines. Notably, it exceeds the runner-up model by 6.93\% and 5.89\% for NDCG@40, and by 7.05\%  and 7.2\%  for NDCG@20 in LastFM and ML-1M datasets, respectively. For larger datasets, our model surpasses the second best models by $8.3\%$ and $2.87\%$ for NDCG@20, and by $4.9\%$ and $3.23\%$ for NDCG@40 for Mind-F and Alibaba-iFashion datasets. The superior performance of the KHGRec framework can be primarily attributed to two key factors: Firstly, it leverages a novel heterogeneous hypergraph for effective signal propagation across the network, encompassing both collaborative and knowledge signals. Secondly, the relational-aware hypergraph attention mechanism helps to exploit the rich semantics extracted from the CKHG, hence boosting the performance and improving explainability.
    
    \item \textbf{Effect of Hypergraph-aware Mechanism:} The comparative analysis of the performance metrics reveals a notable trend: models employing hypergraph structures, such as DHCF, HCCF, and SHT, consistently outperform those using conventional graph structures like LightGCN and SGL.
    This superiority stems from the hypergraph models' enhanced handling of complex user-item interdependencies, which emphasizes the hypergraph approach's efficacy in modeling higher-order structures for more accurate predictions. Among the hypergraph-based models, our KHGRec notably exceeds its closest rival, SHT, by margins of 7.05\% and 6.93\% for NDCG@20 and NDCG@40 in the LastFM and ML-1M datasets, respectively. For more large-scale datasets, KHGRec outperforms the best-performing hypergraph-based methods by 4.9\% and 9.7\% for NDCG@40 on news and fashion datasets. This performance boost of KHGRec is largely attributed to the innovative \textit{Self-aware Hypergraph Encoder}, which synergizes with the Transformer-based network's proficient learning of neighboring nodes' importance.
    
    \item \textbf{Effect of KG-enhanced Mechanism:} In our comparative analysis of various baseline models, it becomes evident that those coupled with knowledge graphs (KGs), such as KGAT, KGIN, KGCL, and KGRec, demonstrated superior performance. This finding stresses the value of incorporating supplementary information like knowledge entities to refine the learning process of user preference patterns. Notably, KGRec exhibits robust performance in most scenarios for LastFM, MovieLens, Mind-F, and Alibaba-Fashion datasets, illustrating the benefits of accounting for the semantic structure of triplets in KGs to enhance the expressiveness of user and item embeddings. Despite the efficacy of these KG-enhanced baselines, a significant performance differential is observed between our KHGRec model and its counterparts. This suggests that KHGRec's integration of auxiliary knowledge with high-level topological information contributes substantially to its effectiveness. Specifically, KHGRec surpasses the next best KG-enhanced model, KGRec, by margins of 7.2$\%$ and 12.6\% for NDCG@20 and NDCG@40 on LastFM and Movielens-1M datasets, respectively. For the LastFM dataset, these figures are 7.05\% and 6.93\% for NDCG@20 and NDCG@40 metric, respectively. For Mind-F and Alibaba-iFashion, the NDCG@20 margins between the second-best model are 11\% and 7.5\%, respectively. Differing from KGRec, our approach focuses on both the interaction and knowledge representations of items, synergizing them through an efficient attention mechanism. Moreover, incorporating cross-view contrastive learning fosters a complementary relationship between local and global scope representations, yielding more meaningful and preference-aware expressions.

\end{itemize}

\subsection{Runtime Analysis (RQ2)}
\label{ssec:runtime}
\begin{figure}[!h]
	\centering
     \subfigure{
    	\includegraphics[width=0.9 \linewidth]{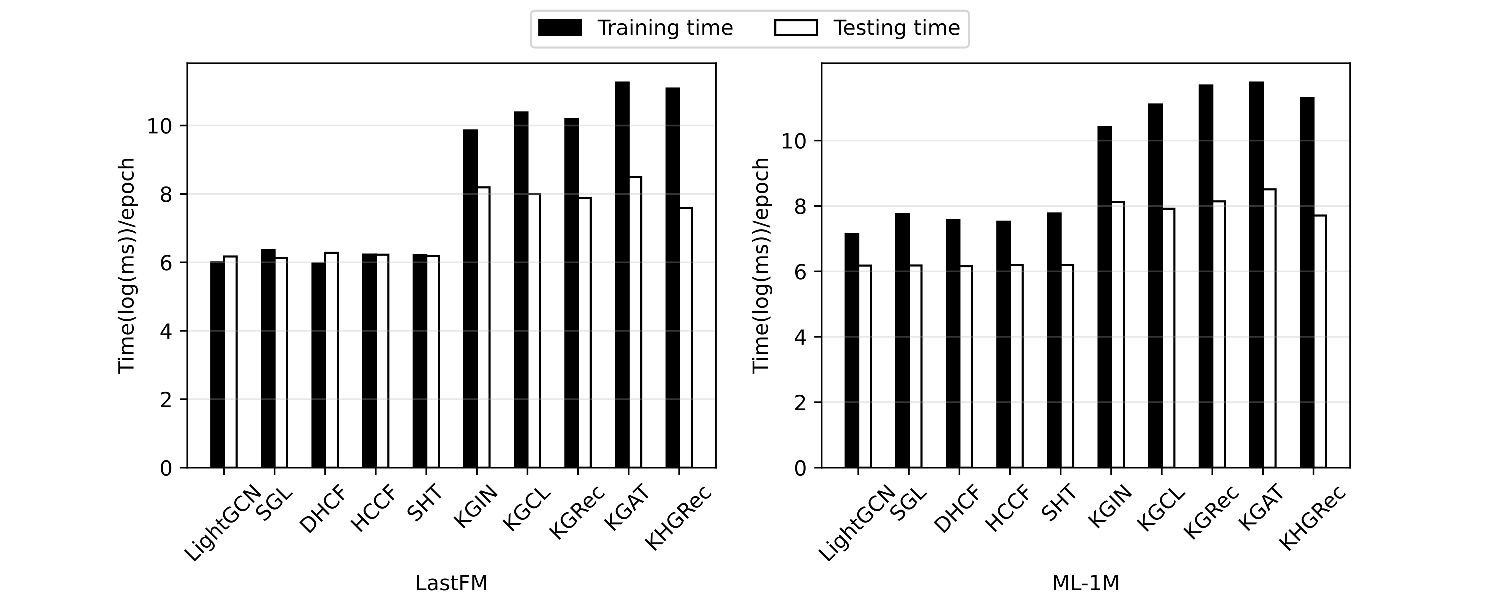}
    }
 \hspace{0.05\textwidth}
    \subfigure{
    	\includegraphics[width=0.9 \linewidth]{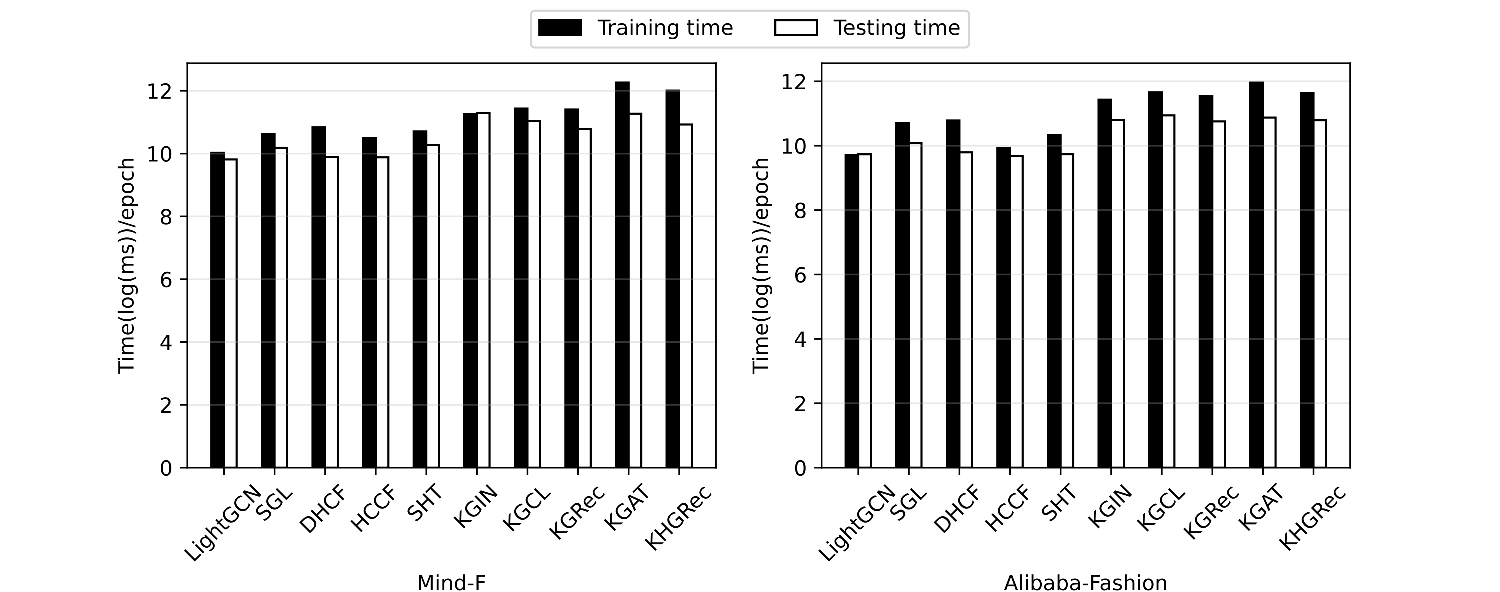}
    }	
    \caption{Runtime comparison of the proposed model and baseline methods.}
    \label{fig:runtime}
\end{figure}

In this section, we measure the running time of our proposed model and compare it with other baseline methods in terms of training time and testing time. The detailed statistics of running time for the proposed and baseline methods are summarized in \autoref{fig:runtime}.
An overarching observation reveals that the training duration of KG-based methods notably exceeds that of CF-based methods. This discrepancy arises from the need for KG-based approaches to integrate user-item interactions and external knowledge from knowledge graphs (KGs) into their embedding learning process, thereby increasing the model's complexity. The empirical analysis indicates that our model consistently ranks second to third in terms of training duration across all datasets, with training times exceeding the average by factors of 3.70, 2.04, 2.21, and 1.63 for LastFM, MovieLens, Mind-F, and Alibaba-Fashion, respectively. This prolonged duration can be attributed to the extensive network size inherent in our model.\\
Similarly, regarding inference time, it is evident that the testing duration of KG-based methods surpasses that of CF-based methods. Our proposed model consistently ranks third to fourth in terms of inference duration across all datasets, with testing times surpassing the average by factors of 1.08, 1.19, 0.94, and 1.23, respectively. Within the KG-based methods category, our model's inference time closely aligns with other methods, ranking first for LastFM and MovieLens datasets, and second for the Mind-F and Alibaba-Fashion datasets regarding inference time efficiency.

\subsection{Cold-start analysis (RQ3)}
\label{ssec:coldstart}
This section examines the cold-start problem in Collaborative Filtering (CF) models, especially for users or items with limited historical interactions. This challenge, common in practical applications, affects CF-based recommendation systems' ability to determine user preferences with sparse historical data. To assess this, we conduct an evaluation focusing on the least interacted users and items within the four datasets, aiming to mirror the scenarios encountered in real-world datasets characterized by limited initial user preference data.

For a detailed analysis, we quantify the interaction frequency of each user. Subsequently, we identify the bottom 10$\%$ of users in terms of interaction frequency, classifying them as cold-start users. For instance, the cold-start user set of the LastFM dataset includes users with fewer than 33 interactions, while the MovieLens dataset encompassed users with fewer than 21 interactions. For the Mind-F and Alibaba-Fashion datasets, these numbers are 45 and 39, respectively. Utilizing these criteria, we then extract a subset of users from the test set that corresponded to these cold-start users. 
This approach allows us to create a test environment that closely simulates real-world conditions, where initial user interaction data is often sparse, thus providing a robust platform to evaluate the efficacy of CF models in mitigating cold-start challenges.

\begin{figure}[!h]
	\centering  
    \subfigure{
    	\includegraphics[width=0.7 \linewidth]{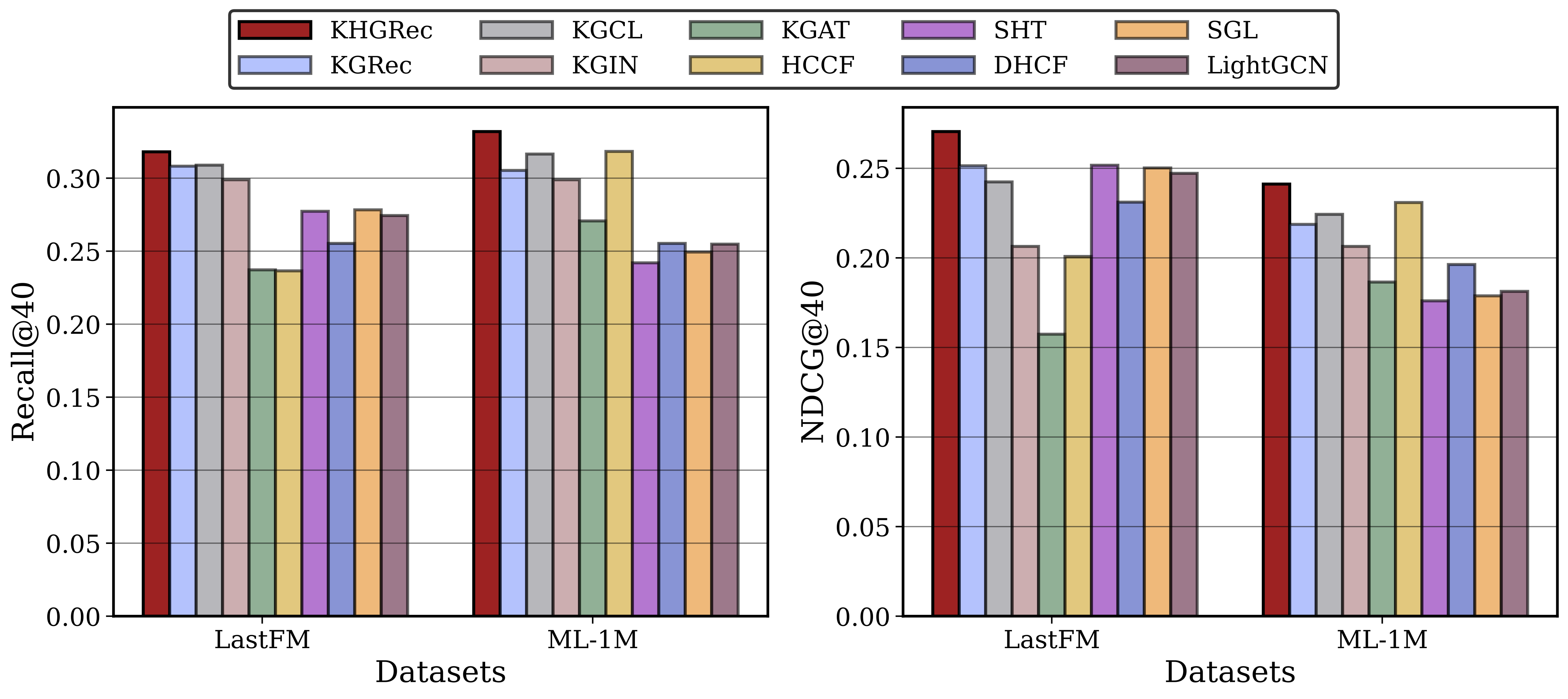}
    }	
	\vspace{-1em}
    \subfigure{
    	\includegraphics[width=0.7 \linewidth]{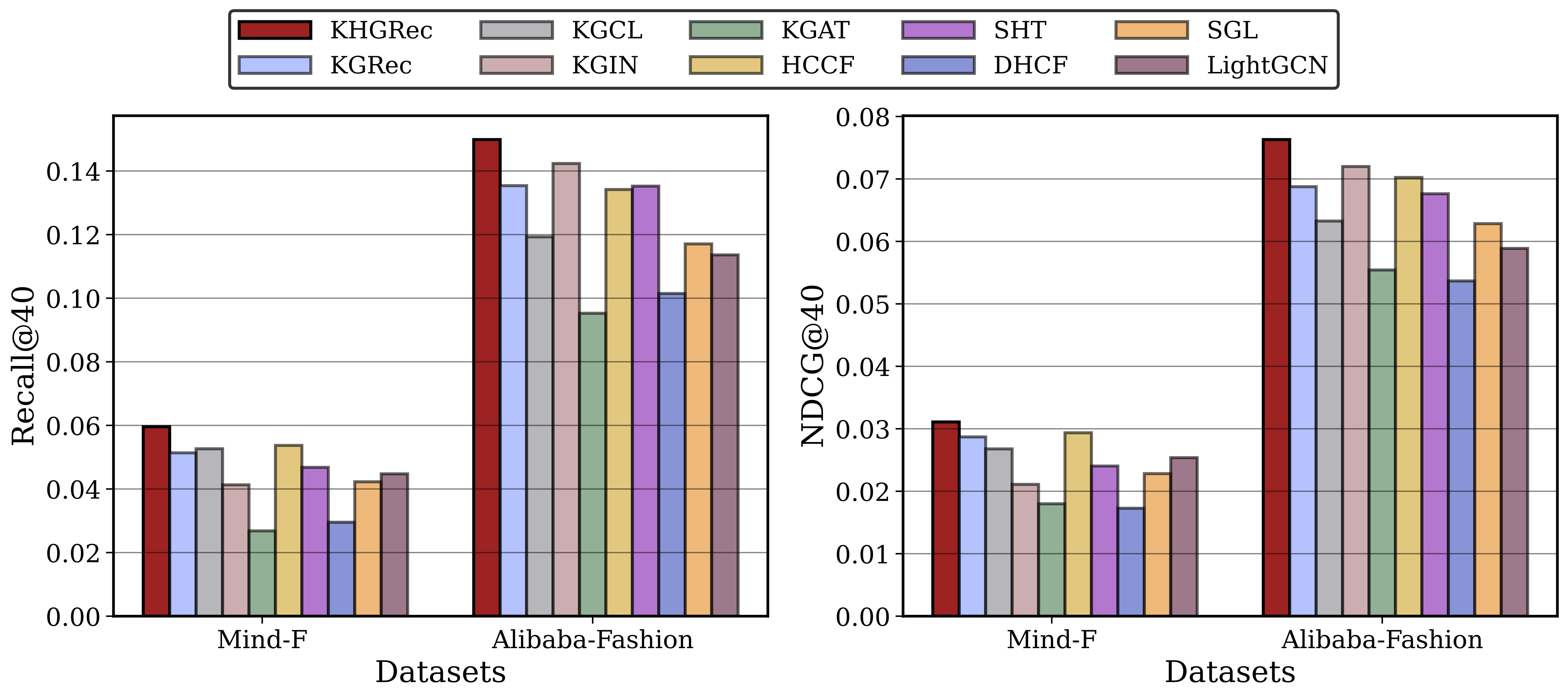}
    }	
    \caption{Comparison on cold-start users.}
    \label{fig:coldstart}
\end{figure}

\autoref{fig:coldstart} presents a comprehensive statistical analysis of performances derived from experiments with the modified datasets. Our findings indicate that KHGRec consistently excels over a range of baseline methods in every experimental setup. Notably, it achieves the most substantial lead over the runner-up model with a 19.8$\%$ and 6.6\% of margins for Recall@20 in the ML-1M and Mind-F, and 12.6$\%$ and 4.2\% of margins for Recall@10 in the LastFM and Alibaba datasets. Furthermore, in comparison to the lowest-performing model, KGAT, KHGRec shows remarkable superiority, surpassing it by 34$\%$ and 71.8$\%$ for Recall@40 and NDCG@40 in the LastFM dataset, respectively. In the context of the ML-1M dataset, KHGRec leads over the least effective model, SHT, by margins of 37.1$\%$ for Recall@40 and 37$\%$ for NDCG@40. For the Mind-F, the performance gap of NDCG@40 between our model and the lowest-performing model is 73.1\% and for the Alibaba dataset, it is 42.3\%. This overall analysis highlights KHGRec's enhanced capability in handling cold-start users compared to the existing state-of-the-art baselines, which can be justified by the ability to capture the complex interdependencies and learn the distinctive yet generalized embedding of the Transformer-based model. 

\subsection{Analysis of noise-resilient capability (RQ4)}
\label{ssec:noiseresilient}
This section studies the capacity of the models to mitigate the noise effect, such as falsely connected user-item pairs, which is prevalent in real-world scenarios. To perform the investigation, we inject random noises into the datasets by randomly creating connections between users and items that the user has not been rated or has not shown interest in before. More specifically, for in-depth analysis, we have applied five different portions of noise injection ranging from 10$\%$ to 50$\%$.

\begin{figure}[!h]
	\centering  
    \subfigure{
    	\includegraphics[width=0.7 \linewidth]{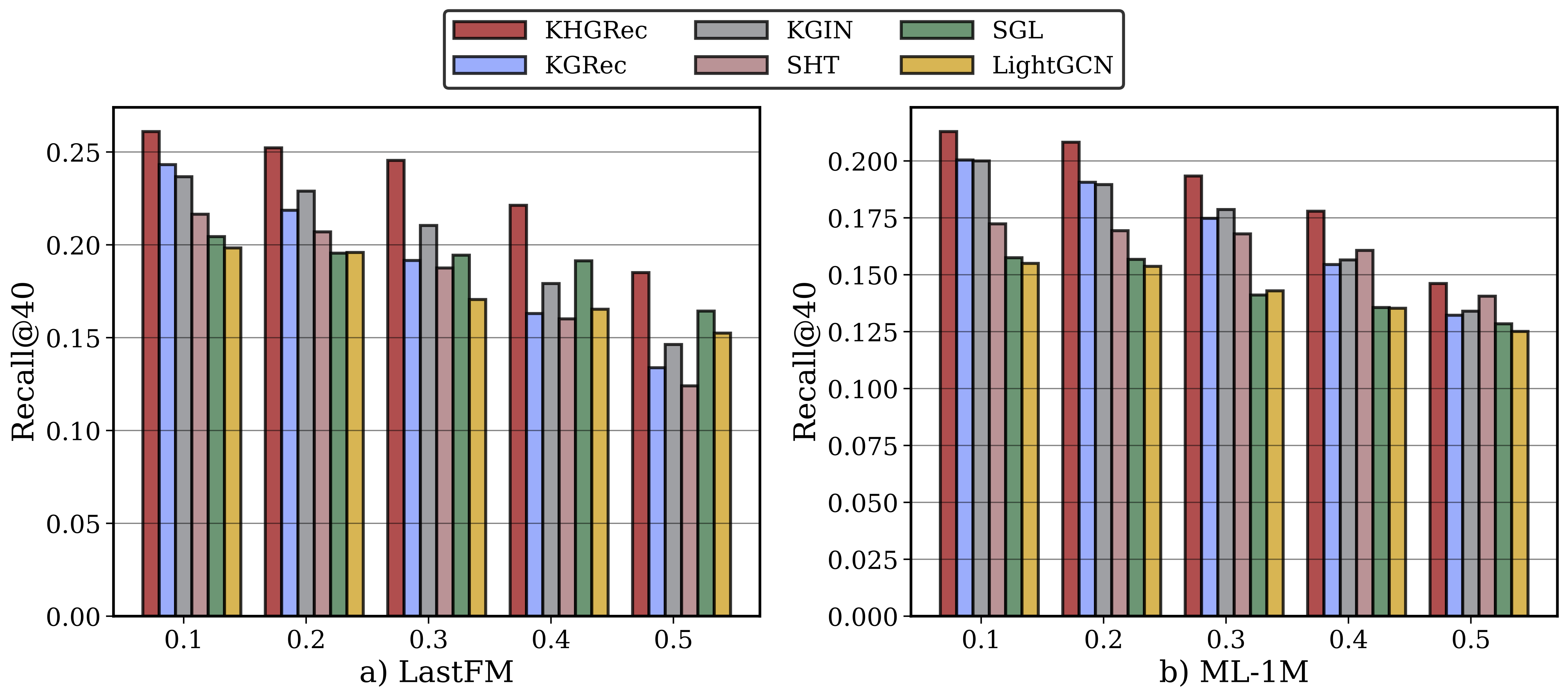}
    }	
	\vspace{-1em}
    \subfigure{
    	\includegraphics[width=0.7 \linewidth]{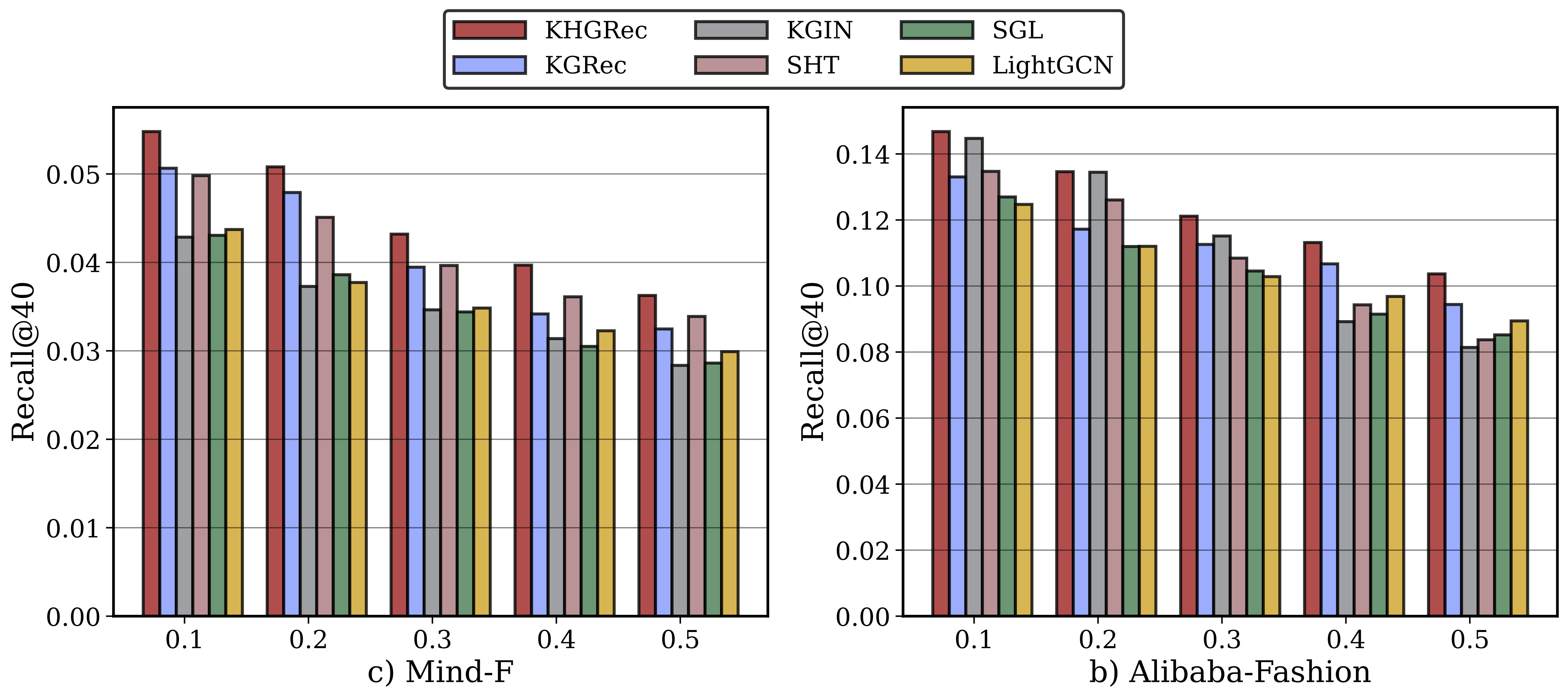}
    }	
    \caption{Comparison of noise-resistant capability.}
    \label{fig:noise_resilient}
\end{figure}

\autoref{fig:noise_resilient} reports a detailed comparative analysis across various noise levels. This analysis reveals a significant decline in the performance of baseline models as the input user-item bipartite graph incorporates increasing amounts of invalid preference data. Notably, the runner-up model, KGRec, shows a marked decrease in Recall@40 accuracy at noise levels of 10\% and 50\%, with reductions surpassing 60\% in the music dataset, 50\% in the movie dataset, 55.9\% for the news dataset, and 41.4\% for the fashion dataset, respectively.

On the other hand, our KHGRec model demonstrates a more resilient performance in the face of noise. Specifically, it shows a comparatively modest decline in Recall@40 accuracy at noise levels of 10\% and 50\%, with reductions of about 41$\%$, 45.6$\%$, 51.1\%, and 33.4\% for the LastFM, MovieLens, Mind-F, and Alibaba-Fashion datasets, respectively. Furthermore, it is noteworthy that KHGRec consistently achieves the highest performance in all test settings, with a notable 7.4$\%$ accuracy gap over the runner-up model, KGRec, at a 50$\%$ noise level. This indicates KHGRec can effectively mitigate noise impact through the encoding of representations enriched with multiple signals sourced from diverse graphs and their intricate group-wise topological information. This superior performance is attributed to the Transformer network's enhanced generalization capabilities, enabling effective learning with limited training data.

\subsection{Analysis of effects of training data proportion (RQ5)}
\label{ssec:trainingproportion}

\begin{figure}[!h]
	\centering  
    \subfigure{
    	\includegraphics[width=0.7 \linewidth]{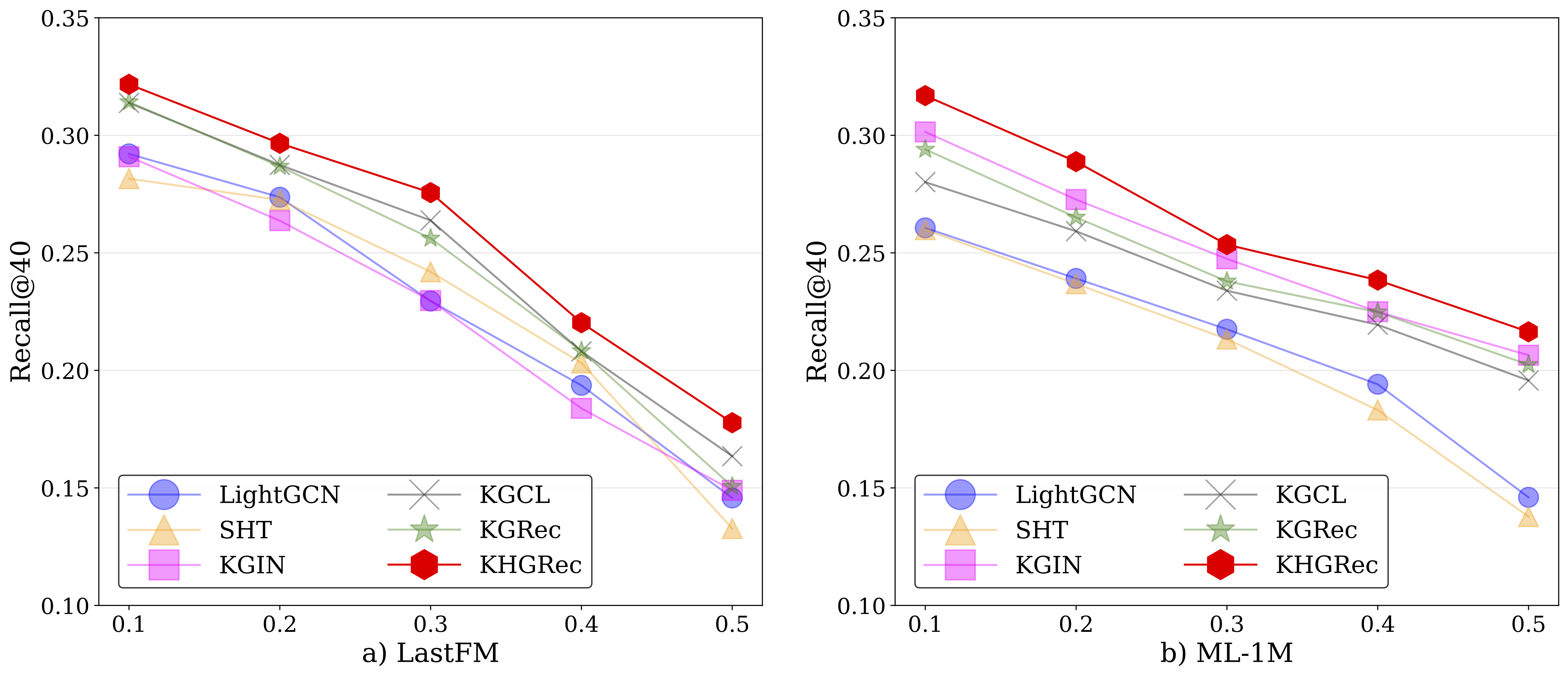}
    }	
	\vspace{-1em}
    \subfigure{
    	\includegraphics[width=0.7 \linewidth]{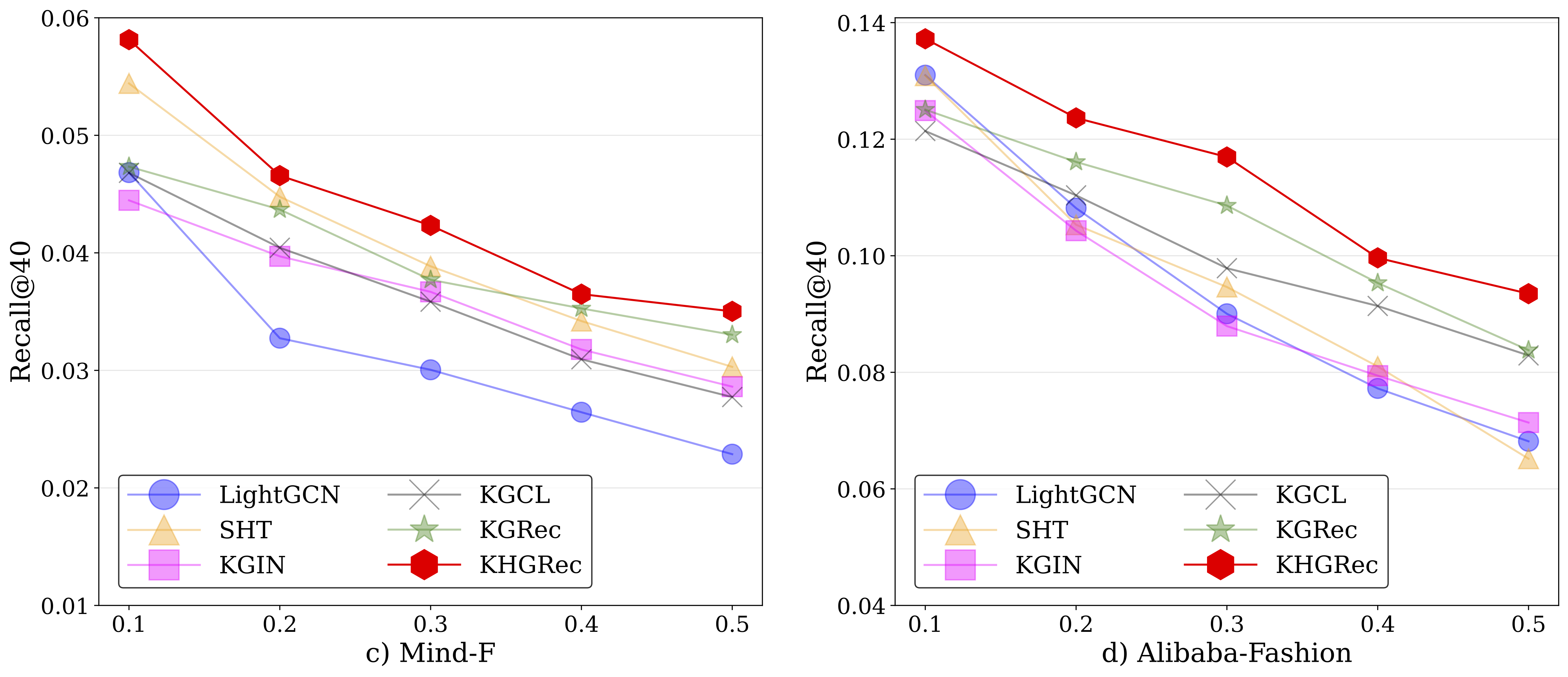}
    }	
    \caption{\textbf{Analysis of the effect of training data percentage removed.}}
    \label{fig:missingdata_exp}
\end{figure}

This section examines how different proportions of training data affect the performance of various models. The absence of adequate training data can intensify model bias due to imbalances in the dataset, impeding the ability of models to discern significant patterns in user-item interactions, consequently leading to inferior recommendation quality. We conduct experiments by randomly omitting 10$\%$, 20$\%$ 30$\%$ 40$\%$, and 50$\%$ of the training data from the total dataset. \autoref{fig:missingdata_exp} illustrates the comparative performance of our proposed KHGRec model against various baseline models under each experimental condition.

Overall, KHGRec consistently outperforms across all test scenarios. Specifically, in the LastFM dataset, KHGRec surpasses the next-best model, KGCL, by an average of 4.45\%. In the ML-1M dataset, it exceeds the runner-up model, KGIN, by 4.07\%. Additionally, KHGRec significantly outperforms the lowest-performing model, LightGCN, by margins of 13.83\% and 22.18\% in LastFM and ML-1M, respectively. For the Mind-F and Alibaba datasets, we can observe 42.4\% and 27.6\% performance decreases for Recall@40, respectively.

This improved performance can be credited to the Transformer network's capability for improving generalization, allowing for learning a well-performed model even with limited training data.

\subsection{Ablation Study of KHGRec Framework (RQ6)} 
\label{ssec:exp_ablation}
In this study, we discuss how significantly each component in KHGRec contributes to the performance gain we observed in the foregoing section. To investigate the importance of each module, we create four variants by discarding four different types of main components: Cross-view contrastive learning, Hypergraph Encoder, Global encoder, and Attention feature fusion module, respectively. The overall results of the study are reported in Table \ref{tab:ablation}.

\begin{itemize}
    \item \textbf{Effect of Cross-view Contrastive Learning} 
    We first assess the impact of cross-view contrastive learning on the performance of our model. The absence of contrastive supervision between local and global representations leads to a notable decrease in both Recall@40 and NDCG@40 metrics when compared to the performances achieved by the original KHGRec model. Specifically, for the Recall@40 metric of all four datasets, the absence of this feature results in performance declines of 13.8$\%$, 4.6$\%$, 1.34$\%$, 7.1$\%$, respectively. Analogously, for the NDCG@40 metric, there observed performance gaps of 24.1$\%$, 8$\%$, 6.01\%, and 10.92\% in each dataset, respectively.
    These findings highlight the significant role that contrastive learning plays in the KHGRec framework. By facilitating the effective capture of both explicit and implicit signals, contrastive learning ensures the generation of more distinct representations of items and users. This approach is crucial for enhancing the model's ability to learn the distinctive features of various user and item profiles, thereby contributing to its overall performance and accuracy.
       
    \item \textbf{Effect of Hypergraph Encoder} 
    We conduct an experiment to evaluate the impact of the Hypergraph Encoder within our model. Specifically, in this experiment, we replace the hypergraph convolution with the simple graph convolution introduced in~\cite{he2020lightgcn}. The results, as presented in \autoref{tab:ablation}, reveal a marked drop in the model using the simple graph convolution. Specifically, this variation leads to a decrease of 7.76$\%$ in Recall@40 for the LastFM dataset, 6.5$\%$ for the ML-1M dataset, 8.18\% for Mind-F dataset, and 3.52\% for Alibaba-Fashion dataset. Additionally, in terms of the NDCG@40 metric, the margins observed are 9.19$\%$ for LastFM, 8.03$\%$ for ML-1M, 11.63\% for Mind-F, and 5.78\% for Alibaba-Fashion, respectively. These findings strongly affirm the importance of Hypergraph Encoder in the model's architecture. Specifically, its ability to encode complex group-wise interdependencies among users and items significantly enhances the model's predictive accuracy and recommendation quality. 
    
    \item \textbf{Effect of Global Encoder} We also evaluate the impact of excluding the \textit{Global Relation-aware Hypergraph Encoder} from KHGRec. The results, as detailed in Table.\ref{tab:ablation}, indicate a notable performance discrepancy between this modified version and the original model configuration, particularly in the context of the LastFM dataset. The absence of this component results in a substantial margin of 18.7$\%$, 8.04$\%$, 14.76\%, and 7.3\% in Recall@40 metric and 18.9$\%$, 10.1$\%$, 20.63\%, and 10.88\% in the NDCG@40 metric for each dataset, respectively. The marked decline observed in the modified model demonstrates the significant contribution of the relation-aware hypergraph attention mechanism, which can help to capture and utilize the complex relational dynamics to improve the RecSys's prediction accuracy.

    \item \textbf{Effect of Attention Feature Fusion} 
    The removal of the attention feature fusion module from KHGRec leads to a notable decline in performance across various test scenarios. This is particularly evident in the NDCG@40 metric, where the absence of this module resulted in performance drops of 25.4$\%$, 11.7$\%$, 18.95\%, 16.45\% for each music, movie, news, and fashion dataset, respectively. The results from this assessment firmly establish the attention fusion module as an important component of KHGRec, significantly contributing to its ability to deliver enhanced and accurate recommendations. Furthermore, this outcome validates the module's underlying rationale, emphasizing the necessity of differentially weighing dual-channel-based embeddings to improve the model's recommendations. 

\end{itemize}

\begin{table}[htb!]
\centering
\scriptsize
\caption{\textbf{Ablation results of KHGRec with different variants.}}
\label{tab:ablation}
\begin{tabular}{@{}lcc@{\hskip 0.5cm}cc@{\hskip 0.5cm}cc@{\hskip 0.5cm}cc@{\hskip 0.5cm}cc@{}}
\toprule
\multirow{2}{*}{Ablation Settings} & \multicolumn{2}{c}{LastFM} & \multicolumn{2}{c}{MovieLens} & \multicolumn{2}{c}{Mind-F} & \multicolumn{2}{c}{Alibaba-iFashion} \\
\cmidrule{2-3} \cmidrule{4-5} \cmidrule{6-7} \cmidrule{8-9}
& Recall@40 & NDCG@40 & Recall@40 & NDCG@40 & Recall@40 & NDCG@40 & Recall@40 & NDCG@40 \\
\midrule
\text{KHGRec} & 0.3551 & 0.3302 & 0.3289 & 0.3968 & 0.0640 & 0.0334 & 0.1575 & 0.0802 \\
\midrule
\text{w/o CCL} & 0.3118 & 0.2660 & 0.3142 & 0.3674 & 0.0632 & 0.0314 & 0.1463 & 0.0714 \\
\text{w/o Hyper} & 0.3295 & 0.3024 & 0.3088 & 0.3673 & 0.0584 & 0.0295 & 0.1520 & 0.0755 \\
\text{w/o Global Encoder} & 0.2991 & 0.2776 & 0.3044 & 0.3601 & 0.0546 & 0.0265 & 0.1461 & 0.0721 \\
\text{w/o Attention} & 0.3058 & 0.2633 & 0.2932 & 0.3552 & 0.0551 & 0.0271 & 0.1436 & 0.0676 \\
\bottomrule
\end{tabular}
\end{table}

\subsection{Benefits of KHGRec in Improving Explainability (RQ7) }
\label{ssec:exp_qualitative}
\begin{figure*}[!h]
    \centering
    \subfigure{
    \includegraphics[width=0.3\textwidth]{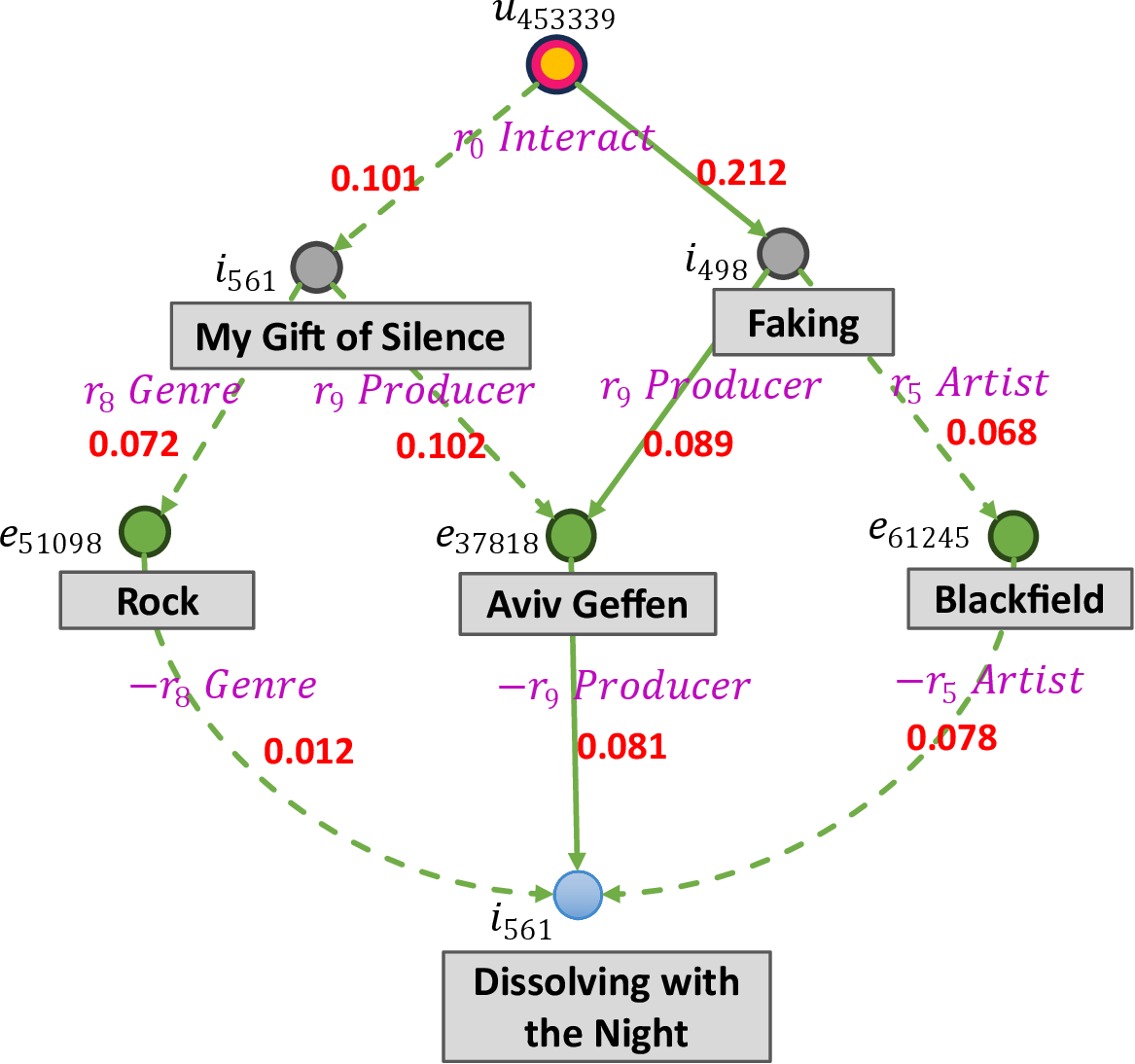}
    }
    \subfigure{
    \includegraphics[width=0.35\textwidth]{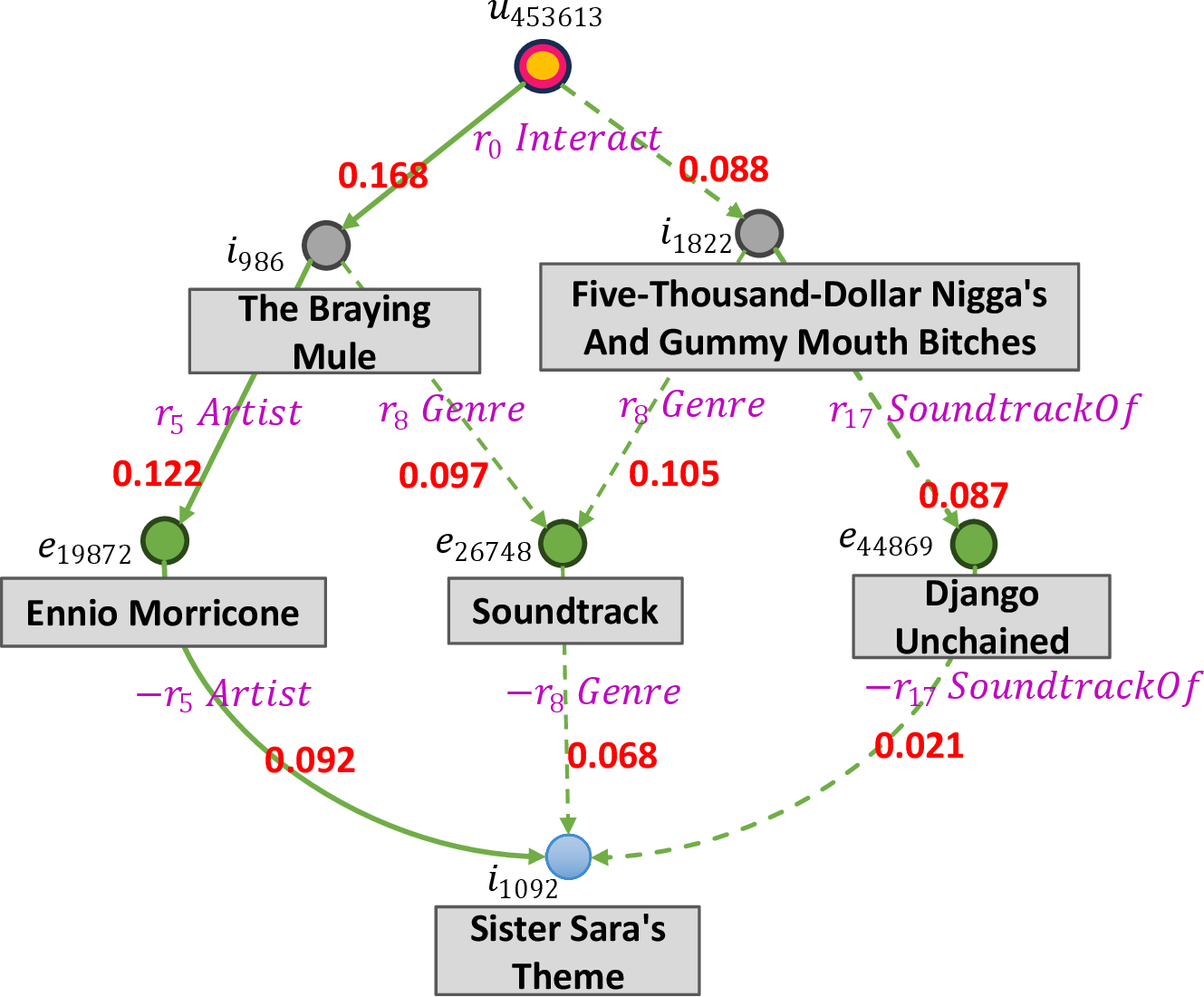}
    }
    \caption{\textbf{Real example from LastFM.}}
    \label{fig:explainability}
\end{figure*}

Thanks to the attention architecture, we justify that incorporation of the wide range dependency can predict more accurate latent user interest in items based on rich semantics. Toward this end, we demonstrate two cases to explain the desired characteristics; with each case, we randomly select one user from the LastFM dataset and his recommended items in the test set. Given the user, we distill two subgraphs shown in Figure~\ref{fig:explainability}, where each connection is assigned a specific attention weight that reflects the importance of that specific connection. We then emphasize the connections with higher attention scores using solid lines. As shown in the first graph in Figure~\ref{fig:explainability}, the solid lines construct a path $u_{453339}$ $\xrightarrow{r_0}$ \textit{Faking} $\xrightarrow{r_{9}}$ \textit{Aviv Geffen}  $\xrightarrow{-r_{9}}$ \textit{Dissolving with the Night}.  
From the observed attention-based path, we can conclude that the song \textit{Dissolving with the Night} is recommended since the user has watched \textit{Faking}, which was also produced by \textit{Aviv Geffen}.
In the second case, the path with highest attention score, $u_{453613}$ $\xrightarrow{r_0}$ \textit{The Braying Mule} $\xrightarrow{r_{5}}$ \textit{Ennio Morricone}  $\xrightarrow{-r_{5}}$ \textit{Sister Sara's Theme}, generates the explanation as follows: The movie \textit{Sister Sara's Theme} is recommended since the user has also watched \textit{The Braying Mule}, which was also performed by \textit{Ennio Morricone}. The examples prove the explanatory power of KHGRec, which also benefits the inference of user interests in items. 

\subsection{Hyperparameter sensitivity (RQ8)}
\label{ssec:exp_sensitivity}
\begin{figure*}[!ht]
\vspace{-1em}
    \centering
    \subfigure{
    \includegraphics[width=0.22\textwidth]{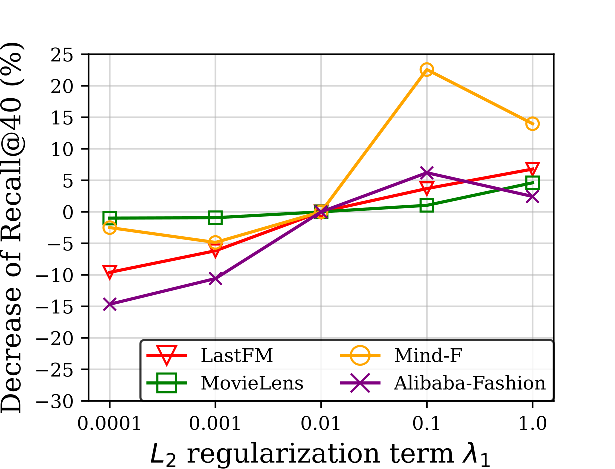}
    }
    \subfigure{
    \includegraphics[width=0.22\textwidth]{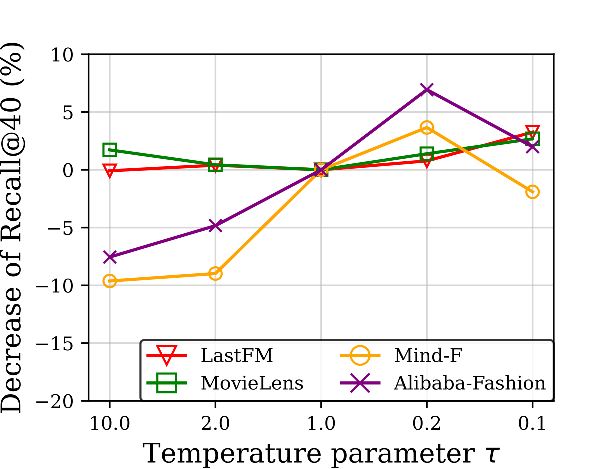}
    }
    \subfigure{
    \includegraphics[width=0.22\textwidth]{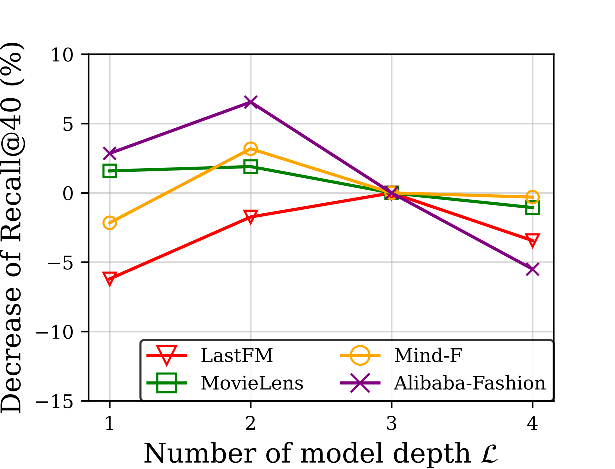}
    }
    \subfigure{
    \includegraphics[width=0.22\textwidth]{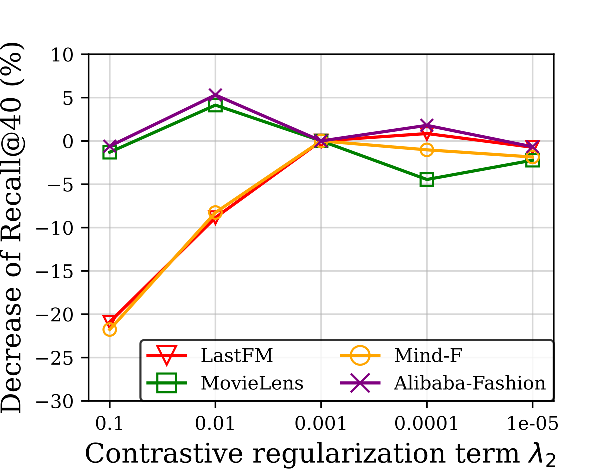}
    }
    \subfigure{
    \includegraphics[width=0.22\textwidth]{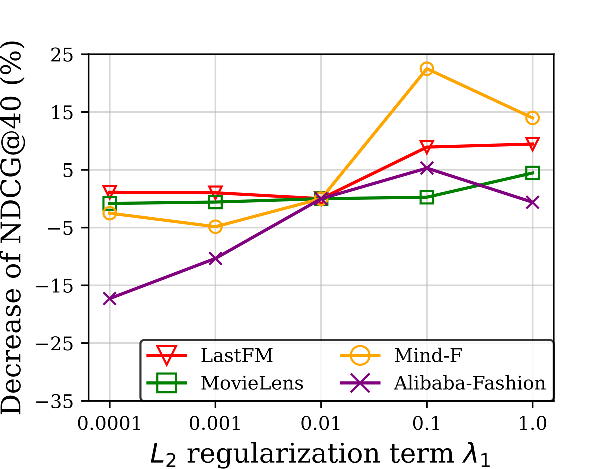}
    }
    \subfigure{
    \includegraphics[width=0.22\textwidth]{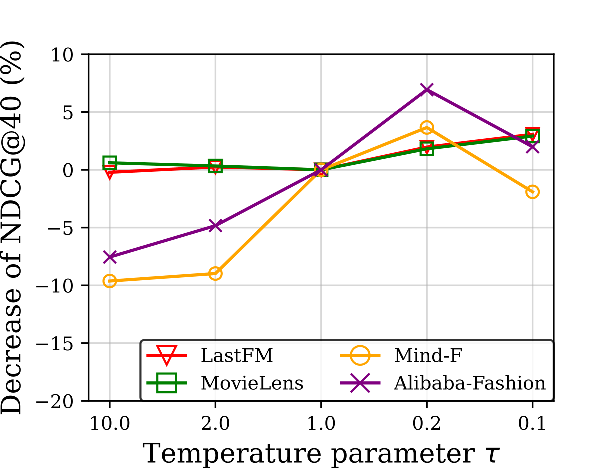}
    }
    \subfigure{
    \includegraphics[width=0.22\textwidth]{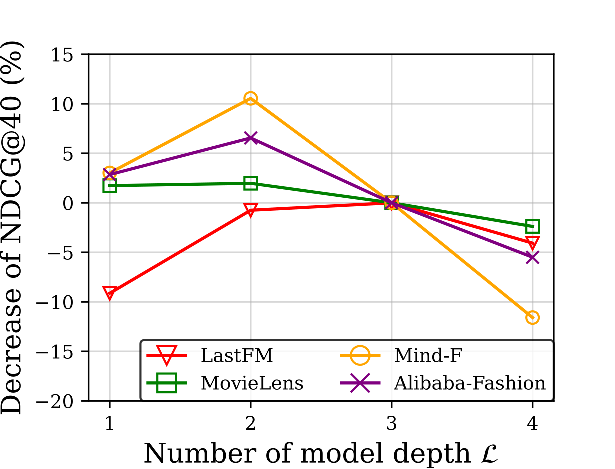}
    }
    \subfigure{
    \includegraphics[width=0.22\textwidth]{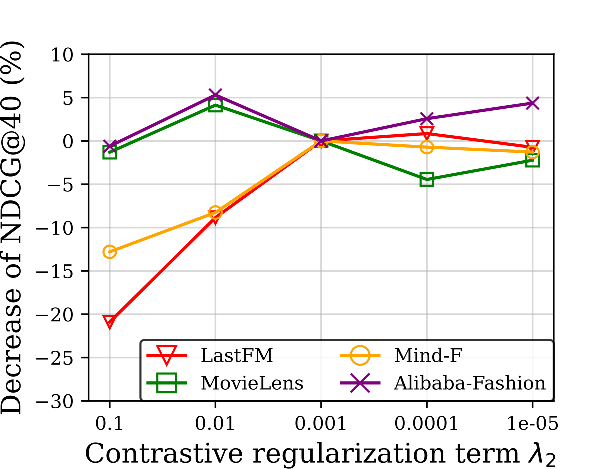}
    }
    \caption{\textbf{Hyperparameter study of the KHGRec.}}
    \label{fig:sensitive}
    \vspace{-1em}
\end{figure*}

In this study, we investigate the sensitivity of KHGRec to changes in key hyperparameters, including $L_2$ regularization term, temperature for CL graph augmentation $\tau$, model depth $L$, regularization term for CL $\lambda_2$. The detailed statistics are shown in Figure~\ref{fig:sensitive}.

\sstitle{Effect of L2 regularization term}
To assess the impact of the $L_2$ regularization term on model performance, we conduct an experiment by varying $L_2$ from 0.0001 to 1, while keeping other hyperparameters constant. The results unveil the trend that as the weight decay parameter increases, there is a corresponding enhancement in both metrics for both datasets. Specifically, for the music dataset, the performance, particularly in terms of NDCG at various k values (e.g., 20, 40), shows a significant improvement when $L_2$ values exceed 0.01. On the other hand, the performance for the movie dataset exhibits a more gradual improvement, steadily increasing with $L_2$ values exceeding 0.01.
Alibaba-iFashion dataset shows an analogous pattern to LastFM. However, performance decreases significantly when the regularization term transcends 0.1. Lastly, in the Mind-F dataset, the recall@40 value rapidly increases when the term reaches 0.1, and the performance diminishes substantially when it is set as 1.

\sstitle{Effect of the temperature for CL graph augmentation}
In our study, we also explore the influence of varying the temperature parameter $\tau$ in the contrastive loss function. This parameter is varied across a range of values: $\{10, 2, 1, 0.2, 0.1\}$, to assess its impact on model dependency and performance. Our findings indicate that for the ML-1M dataset, optimal Recall@k and NDCG@k scores are achieved when $\tau$ is set to 0.1, with the lowest scores observed at a $\tau$ value of 1. More precisely, a notable improvement of 3.7$\%$ in Recall@20 is recorded when $\tau$ is adjusted from 1 to 0.1 for the movie dataset. For the music dataset, the model's lowest performance for the Recall@20 metric is achieved when $\tau$ is set to 0.2. However, the overall performance difference between the best and worst scenarios is relatively minimal, suggesting that KHGRec's performance is not significantly impacted by the temperature parameter for the music and movie datasets.
On the other hand, Mind-F and Alibaba-Fashion datasets are heavily influenced by the temperature term, where the NDCG@40 margins between the best configuration and the least favorable setup are 14.7\% and 15.6\%. Hence, it is recommended that the desirable temperature term be carefully selected for the optimal result on large datasets.

\sstitle{Effect of Model Depth $L$} 
We vary the depth of the KHGRec model, denoted as $L$, to evaluate the impact of utilizing multiple embedding propagation layers. Specifically, we experiment with different numbers of layers, ranging from 1 to 4. For the LastFM dataset, optimal performance in both Recall and NDCG metrics at k values (i.e., 20, 40) is achieved with three layers. However, a significant drop in performance is seen as the number of layers continued to grow. Quantitatively, the performance difference between the highest and lowest results is observed to be 12.09$\%$ for NDCG@20 and 10.04$\%$ for NDCG@40. Conversely, in the MovieLens dataset, the peak performance is achieved with two layers. The performance gap between the best and worst outcomes for ML-1M is 5.1$\%$ for NDCG@20 and 4.4$\%$ for NDCG@40, respectively.
A comparable result is observed for the news and fashion datasets, both of which demonstrate peak performance when the number of layers is set to 2. Notably, Recall@40 for Alibaba-iFashion and NDCG@40 for Mind-F datasets significantly drop when the number of layers is larger than 2.

\sstitle{Effect of the regularization term for contrastive loss function} 
Finally, we examine the influence of the contrastive regularization term, $\lambda_2$, which governs the significance of the contrastive loss relative to other loss functions within the model. Overall, the model reaches the peak performance at different $\lambda_2$ for each dataset, which is 0.01 for the ML-1M dataset and $1e^{-4}$ for the LastFM dataset. Specifically. for the LastFM dataset, the disparity in performance between the optimal and least effective settings for Recall@20 and Recall@40 is 31.5$\%$ and 27.5$\%$, respectively. Mind-F dataset also demonstrates a similar trend, where the disparity of Recall@40 between optimal and worst configuration is 27.8\%. This significant variance indicates that LastFM's performance is heavily dependent on $\lambda_2$. In contrast, the impact of $\lambda_2$ on the ML-1M dataset is comparatively less pronounced. The performance gap between the best and worst outcomes for Recall@20 and Recall@40 was measured at 7.7$\%$ and 8.9$\%$, respectively. Analogously, the influence of the term on the Alibaba-Fashion dataset is nuanced and exhibits the best performance when the term is set to 0.01.

\section{Related Work}
\label{sec:related}

\subsection{Model-based Collaborative Filtering Learning}

To enable the automatic generation of suitable recommendations, a range of research has put their attention to model-based Collaborative Filtering(CF) algorithms which involves building a predictive model that can predict user preferences on a list of items. With the historical information of interaction between user and items(e.g., click and view history, user's ratings on items), model-based CF methods aim to capture the latent interaction patterns and parameterize the users and items in a shared vector space. Based on the closeness of users and items in a latent space, the learned model suggests the items that the target user might like~\cite{nguyen2023poisoning,nguyen2023example,nguyen2014reconciling,nguyen2015smart,thang2015evaluation}.
Conventionally model-based CF can be Matrix Factorization\cite{koren2009matrix}, so-called MF, and it decomposes the user and item interaction matrix into two-lower dimensional user and item matrices. The main goal of MF is to enforce the product of these two matrices to approximate the original interaction matrix. Despite its simplicity and scalability, it struggles to capture explicit collaborative signals and non-linear intricate relationships, thus leading to the lack of expressiveness of embeddings, which hinders the potentially related user and items from being closely positioned in a vector space.

\subsection{Graph-based Collaborative Filtering Learning}

To further encode explicit collaborative signals, many recent works have investigated constructing a bipartite graph that reflects pairwise dependencies between users and items.
One of the convention works such as ItemRank\cite{gori2007itemrank} utilizes random walks to assign scores to items based on their connectivity within the user-item interaction graph. Another typical method SimRank\cite{jeh2002simrank} measures the topological similarity between nodes in a graph based on their overlapping neighbors. 
Recently, graph neural networks (GNNs)~\cite{thang2022nature,duong2022efficient,tam2021multi,nguyen2023isomorphic,tong2022joint} have opened a new paradigm for incorporating high-hop neighboring information to enrich the expressiveness of the embeddings. Graph Convolutional Matrix Completion(GC-MC)\cite{berg2017graph} combines GCNs and matrix completion method. AGCN\cite{wu2020joint} dynamically assigns weights to user-item interactions and graph connections using the attention mechanism. Mixgcf\cite{huang2021mixgcf} proposes hop mixing method for negative sampling. LightGCN\cite{he2020lightgcn} simplifies the GCN framework by removing redundant components such as nonlinear activation to further make the model CF-oriented. UltraGCN\cite{mao2021ultragcn} even further simplifies the GCN approach for recommendation by eliminating explicit message-passing and adding constraint loss. LightGCL\cite{cai2023lightgcl} proposes an effective simplified contrastive learning framework for contrasting different views of augmented graph signals. Recently, models such as LLMRec\cite{wei2024llmrec} further harnesses large language model(LLM) for reliable de-noised data augmentation. Graph-inspired CF models have achieved significant improvements, however, there exist more complex and higher connectivity levels between users and items in real-world scenarios in which standard graph structure often falls short~\cite{nguyen2015tag,hung2019handling,zhao2021eires,huynh2021network,duong2022deep}. 

\subsection{Hypergraph-aware Recommendation}

Hypergraph, a generalization of a graph in which edges connect to a set of any number of nodes instead of linking pairs of nodes like in a regular graph, has attracted recent studies of its capability to represent informative connectivity information. This sheds light on modeling group relationships which benefits the recommendation systems to gather in-depth collaborative information when a number of collective interactions are available.
DHCF~\cite{ji2020dual} learns the embedding through dual channels and then leverages message-passing methodology on hypergraphs to learn high-order interplays.
HCCF~\cite{xia2022hypergraph} parameterizes the hypergraph structure to model intricate topology with contrastive learning. UPRTH\cite{yang2024unified} utilizes task hypergraphs and a transitional attention layer to generalize pretext tasks to hyperedge prediction. HypAR\cite{jendal2024hypergraphs} leverages two separate modules to provide explainable recommendations based on high-order dependencies. 

Unlike those existing methods, our work aims to encapsulate the heterogeneous hypergraph structure of collaborative signals by borrowing the key idea from transformer architecture for enhanced representation learning. Furthermore, we apply group-wise structuring not only on preference information but substitute knowledge information by introducing collaborative knowledge hypergraph, so that the model can further utilize intricate topological dependencies from given data.

\subsection{Recommendation with Knowledge graph}

Due to the deficiency of accumulated preference history and cold start problems, recommendation tasks often struggle to predict latent interaction accurately. To supplement data sparsity, a line of studies started to leverage Knowledge Graphs(KGs) as additional information by enlarging the scope of side information connected to items, users, and their attributes.
KG-enhanced CF models can be categorized into three main types: embedding-based, path-based, and hybrid models\cite{guo2020survey}.
Embedding-based CF methods \cite{zhang2018learning} vectorize KGs into low dimensional space and estimate the probability whether the user would like the item based on their embeddings.
For example, CFKG\cite{zhang2018learning} includes user information when building KG and involves relation information when calculating the closeness between the target user and items.
Another approach, path-based methods\cite{hu2018leveraging}, extracts the relationship patterns by constructing meta-paths. RippleNet\cite{wang2019exploring} builds ripple set attempts to encapsulate the high-order user interests into representations to provide recommendations. However, group-wise relationships, such as a group of users selecting one particular item, are still ignored. 
MetaKRec\cite{wang2022metakrec} builds Meta-KGs based on the knowledge extracted from both KG and interaction data and incorporates collaborative signals through lightweight GCN layers. EMKR\cite{gao2023enhanced} opts to leverage two channels to learn embeddings from both observed interactions and possible latent interactions. MetaKG\cite{wang2022metakrec} integrates a collaborative-aware meta-learner and a knowledge-aware meta-learner to address cold-start problems.

Despite the advancement of the existing KG-enhanced recommenders, they often overlook group-wise connectivity formed within KGs, which can facilitate explainable and accurate recommendations based on high-order relationships among the entities. To this end, our model goes beyond the current research by pass-forwarding novel graph design, CKHG, through a collaborative knowledge hypergraph encoder. 
Specifically, the encoder aims at capturing the heterogeneous relational information of the constructed CHKG, which has not been considered by the current studies to our best knowledge~\cite{nguyen2022model,nguyen2022detecting,trung2022learning,nguyen2024survey,nguyen2022survey}.

\section{Conclusion}
\label{sec:con}
This paper presents a novel framework for a knowledge-based recommender system, KHGRec, to effectively capture the group-wise characteristics in both the user-item bipartite graph and the knowledge graph. Moreover, we also propose a novel relational-aware hypergraph encoder that leverages the attention mechanism to capture the complex relational-aware dependencies in the knowledge graph. The proposed model achieves state-of-the-art performance on the recommendation task compared to other baselines. Specifically, our proposal achieves 7.2$\%$ of NDCG@20 margin against the runner-up model in the MovieLens dataset. We further validate the superiority of our proposal towards the baselines in being resilient for adversarial conditions such as cold-start, noise influx, and sparse data with significant margins. Moreover, our model also works effectively in scenarios with limited training data, while providing explainable insights to explain the decisions of users. Extensive experiments on several real-world datasets have been conducted to demonstrate the superiority of KHGRec as compared to various state-of-the-art models.
 
In future work, we would like to extend our work to various settings such as trustworthy recommendation~\cite{wang2022trustworthy}, streaming recommendation~\cite{zhao2023mbsrs}, and session-based recommendation~\cite{wang2021survey,ho2022efficient}. Furthermore, we plan to design an effective negative sampling mechanism and augmentation process to address the data sparsity. We will further focus on incorporating recent prosperous techniques such as pre-training and large language models(LLMs) while pertaining the transferability across different datasets and domains.

\vspace{-1em}

%\bibliographystyle{../elsarticle-num-names}
%\bibliography{../ref,../ref_h,../ref_b}

\end{document}